\documentclass[useAMS,usenatbib]{mn2e}

\usepackage{epsfig}
\usepackage{aas_macros}
\usepackage{natbib}
\usepackage{times}
\usepackage{float}

\newbox\grsign \setbox\grsign=\hbox{$>$} \newdimen\grdimen \grdimen=\ht\grsign
\newbox\labox \newbox\gabox \newbox\simpropbox \newbox\wtildebox 
\setbox\gabox=\hbox{\raise.5ex\hbox{$>$}\llap
    {\lower.5ex\hbox{$\sim$}}}\ht1=\grdimen\dp1=0pt
\setbox\labox=\hbox{\raise.5ex\hbox{$<$}\llap
    {\lower.5ex\hbox{$\sim$}}}\ht2=\grdimen\dp2=0pt
\def\ga{\mathrel{\copy\gabox}}
\def\la{\mathrel{\copy\labox}}

\newcommand{\msun}{\mbox{M$_\odot$}}
\newcommand{\cmc}{\mbox{${\rm cm}^{-3}$}}
\newcommand{\yr}{\mbox{${\rm yr}$}}
\newcommand{\myr}{\mbox{${\rm Myr}$}}
\newcommand{\kms}{\mbox{${\rm km}~{\rm s}^{-1}$}}
\newcommand{\kpc}{\mbox{${\rm kpc}$}}
\newcommand{\pc}{\mbox{${\rm pc}$}}
\newcommand{\g}{\mbox{${\rm g}$}}

\newcommand{\be}{\begin{equation}}
\newcommand{\ee}{\end{equation}}
\newcommand{\bea}{\begin{eqnarray}}
\newcommand{\eea}{\end{eqnarray}}

\title{The dynamical evolution of molecular clouds near the Galactic Centre -- I. Orbital structure and evolutionary timeline}
\author{J.~M.~Diederik Kruijssen,$^1$\thanks{kruijssen@mpa-garching.mpg.de}
James~E.~Dale,$^{2,3}$
Steven~N.~Longmore$^4$
\\
$^1$Max-Planck Institut f\"{u}r Astrophysik, Karl-Schwarzschild-Stra\ss e 1, 85748 Garching, Germany\\
$^2$Excellence Cluster `Universe', Boltzmannstra\ss e 2, 85748 Garching, Germany\\
$^3$Universit\"{a}ts-Sternwarte M\"{u}nchen, Scheinerstra\ss e 1, 81679 M\"{u}nchen, Germany\\
$^4$Astrophysics Research Institute, Liverpool John Moores University, IC2, Liverpool Science Park, 146 Brownlow Hill,
Liverpool L3 5RF, UK
}

\begin{document}

\date{Accepted 27 November 2014. Received 6 November 2014; in original form 30 September 2014.}

\pagerange{\pageref{firstpage}--\pageref{lastpage}} \pubyear{2014}
\label{firstpage}

\maketitle

\begin{abstract}
We recently proposed that the star-forming potential of dense molecular clouds in the Central Molecular Zone (CMZ, i.e.~the central few~$100~\pc$) of the Milky Way is intimately linked to their orbital dynamics, potentially giving rise to an absolute-time sequence of star-forming clouds. In this paper, we present an orbital model for the gas stream(s) observed in the CMZ. The model is obtained by integrating orbits in the empirically constrained gravitational potential and represents a good fit ($\chi_{\rm red}^2=2.0$) to the observed position-velocity distribution of dense ($n>{\rm several}~10^3~\cmc$) gas, reproducing all of its key properties. The orbit is also consistent with observational constraints not included in the fitting process, such as the 3D space velocities of Sgr~B2 and the Arches and Quintuplet clusters. It differs from previous, parametric models in several respects: (1) the orbit is open rather than closed due to the extended mass distribution in the CMZ, (2) its orbital velocity ($100$--$200~\kms$) is twice as high as in previous models, and (3) Sgr~A$^*$ coincides with the focus of the (eccentric) orbit rather than being offset. Our orbital solution supports the recently proposed scenario in which the dust ridge between G0.253+0.016 (`the Brick') and Sgr~B2 represents an absolute-time sequence of star-forming clouds, of which the condensation was triggered by the tidal compression during their most recent pericentre passage. We position the clouds on a common timeline and find that their pericentre passages occurred $0.30$--$0.74~\myr$ ago. Given their short free-fall times ($t_{\rm ff}\sim0.34~\myr$), the quiescent cloud G0.253+0.016 and the vigorously star-forming complex Sgr~B2 are separated by a single free-fall time of evolution, implying that star formation proceeds rapidly once collapse has been initiated. We provide the complete orbital solution, as well as several quantitative predictions of our model (e.g.~proper motions and the positions of star formation `hotspots'). The paper is concluded with a discussion of the assumptions and possible caveats, as well as the position of the model in the Galactic context, highlighting its relation to large-scale gas accretion, the dynamics of the bar, the $x_2$ orbital family, and the origin of the Arches and Quintuplet clusters.
\end{abstract}


\section{Introduction} \label{sec:intro}
Star formation is one of the fundamental physical processes driving the baryonic evolution of the cosmos, from reionizing the Universe at very high redshift to regulating galaxy evolution and depositing metals in the interstellar medium (ISM), enabling the development of planetary systems and eventually life. Despite its critical importance, a fundamental physical understanding of star formation has not been achieved \citep{mckee07,kennicutt12,krumholz14}.

Several complicating factors are to blame for our limited understanding of star formation. For instance, star formation is inherently a multi-scale process, of which the physics connecting the different scales are highly complex. A wide range of recent work has attempted to address this problem by connecting the empiricism of galactic star formation relations to the cloud-scale physics of star formation \citep[e.g.][]{bigiel08,heiderman10,lada10,schruba10,gutermuth11,hopkins11,krumholz12a,burkert13,kruijssen14}, but the problem is far from solved. An additional issue is that gas and young stellar populations are generally probed using indirect tracers, the calibration of which has become a very active field of research \citep[see e.g.][]{leroy11,sandstrom13}.

Perhaps most importantly, it has proven extremely difficult to follow the deeply gas-embedded process of star formation in time, from the initial collapse of giant molecular clouds (GMCs) to the emergence of young stellar clusters or associations \citep{dobbs14,longmore14}. Being able to follow the absolute-time evolution of star-forming GMCs would greatly advance our insight into several current problems in star formation. For example, it would aid current efforts to understand the assembly of the stellar initial mass function \citep[IMF; see e.g.][]{bastian10,offner14}, allow us to directly probe the rapidity of star formation and its time evolution \citep{padoan14}, and calibrate gas and star formation tracers on an absolute timeline.

The advent of the Atacama Large Millimeter Array (ALMA) enables the study of dense and deeply embedded molecular gas at the spatial resolution and sensitivity that was previously only accessible at visible wavelengths with the Hubble Space Telescope. For the first time, it will be possible to follow the star formation process from its earliest stages as a function of time, {\it provided that a reference timeline can be identified}. 

We have recently proposed that the Central Molecular Zone (CMZ, i.e.~the central 500~pc) of the Milky Way may host an absolute-time sequence of star cluster progenitor clouds, of which the collapse has been triggered by their tidal compression\footnote{{While it is well-known that the tidal field can compress an object as it approaches pericentre \citep[e.g.][]{mo10}, it is not the only possible compression agent. Geometric convergence can also compress or extend objects on eccentric orbits, sometimes even driving spiral instabilities near galaxy centres \citep[e.g.][]{montenegro99}. In a follow-up paper (Paper~II), we will show that the dominant deformation mechanism for the clouds under consideration here is the tidal field.}} during a close passage to the bottom of the Galactic potential well near Sgr~A$^*$ \citep{longmore13b}. This picture is supported by a monotonic increase of the star formation activity along the direction of motion, as well as strong indications that the clouds have recently passed pericentre. It is certainly a tempting idea -- an evolutionary sequence of protocluster clouds with a common zero point would greatly aid current efforts aiming to quantify the time evolution of the star formation process.

The CMZ contains a large reservoir of dense molecular gas \citep[$M_{\rm gas}\sim5\times10^7~\msun$,][]{morris96,ferriere07} with properties widely different from the ISM in the Galactic disc. The molecular gas volume density is two orders of magnitude higher than in the disc \citep[$n_{\rm CMZ}\sim10^4~\cmc$ as opposed to $n_{\rm disc}\sim10^2~\cmc$, see e.g.][]{longmore13}, the medium is highly turbulent, with Mach numbers up to ${\cal M}_{\rm CMZ}\sim 30$ \citep{bally88,kruijssen14b}, and the molecular gas temperature is substantially higher than in the disc as well ($T_{\rm CMZ}=50$--$400$~K versus $T_{\rm disc}=10$--$20$~K, see \citealt{ao13} and \citealt{mills13}). The ISM conditions in the CMZ are very similar to those seen in high-redshift galaxies \citep{kruijssen13b}, which have similarly high molecular gas volume densities, turbulent pressures and temperatures \citep[e.g.][]{swinbank11,danielson13}. This implies that a detailed understanding of star formation in the CMZ may actually provide insight into star formation in extreme environments across cosmic time -- in particular at the peak of the cosmic star formation history at redshift $z=2$--$3$ \citep[e.g.][]{madau96,hopkins06b}.

The ISM of the CMZ is well-studied in several recent Galactic plane surveys of high-density gas tracers \citep[e.g.][]{bally10,walsh11,jones12,jackson13}, providing a wealth of observational data to infer the orbital structure of the gas in the CMZ and test the hypothesis of \citet{longmore13b} that the clouds follow an absolute-time sequence of star formation. The first cloud in the proposed sequence is G0.253+0.016 (also known as `the Brick'), which is thought to be the progenitor of a young massive cluster \citep{longmore12} and is extremely well-studied across a wide range of molecular line and continuum observations \citep{lis98,lis01,bally10,kauffmann13,rathborne14,rathborne14b,johnston14}. With such a wealth of observational data of CMZ clouds \citep[also see][]{immer12b,kendrew13,walker14}, a theoretical census of GMC dynamical evolution and star formation in the CMZ is both urgently needed and within reach.

There is a long history of work aimed at constraining the orbital dynamics of GMCs in the CMZ \citep[e.g.][]{binney91,sofue95,englmaier99,sawada04,stark04,rodriguezfernandez08}. However, none of these studies have been able to exploit the recent flurry of high-resolution surveys of high-density gas in the CMZ, which sketch a much clearer picture of the gas dynamics than previous surveys of the more diffuse H{\sc I} and $^{13}$CO(1--0) lines \citep[e.g.][]{burton78,bally87}.

In a series of papers, we aim to address the hypothesis of \citet{longmore13b} in more detail. In this first paper, we combine the recent observational data with orbital modelling to constrain the orbital structure of the dense ($n>{\rm several}~10^3~\cmc$) gas in the CMZ. We show that the sequence of protocluster clouds identified by \citet{longmore13b} follows a coherent structure in position-velocity space. We highlight the many successes of the currently standard, parametric orbital model of \citet{molinari11} in describing the position-velocity structure of the gas, as well as several areas of improvement. By fitting an orbital model to the gas in position-velocity space, we determine where on the absolute timeline the GMCs in the CMZ are situated, allowing us to draw a number of preliminary conclusions regarding the evolution of these clouds and the physics of star formation. In a companion paper (Paper~II), we present numerical simulations of collapsing gas clouds that follow the best-fitting orbit and we answer the question whether the sequence of GMCs in the CMZ indeed represents an absolute timeline.

In \S\ref{sec:obs}, we first discuss the observed kinematics of the molecular gas in the CMZ, present a systematic survey of its position-velocity structure, and list the strengths and weaknesses of the most recent model for the orbital structure of the dense gas in the CMZ. In \S\ref{sec:orbit}, we introduce a new dynamical model that accurately describes the large-scale motion of GMCs in the CMZ. In \S\ref{sec:impl}, we discuss the implications of our model for GMC evolution and the physics of star formation in the CMZ. We also make a number of quantitative predictions for further observational tests of the model. In \S\ref{sec:disc} we summarize our work, discuss the strengths and weaknesses of our model as well as several open questions, and present a brief outlook. The adopted gravitational potential is discussed and validated in Appendix~\ref{sec:potential}, the dependence of our orbital model on the orbital parameters is presented in Appendix~\ref{sec:appendix}, and the complete orbital solution is tabulated in Appendix~\ref{sec:fullorbit} -- a machine-readable table with time steps of $\Delta t=0.01~\myr$ is available in the Supporting Information accompanying this paper.

Throughout this paper, we adopt a mean molecular weight of $\mu=2.3$, implying a mean particle mass of $\mu m_{\rm H}=3.9\times10^{-24}~\g$, and we assume a distance to the Galactic Centre of $R=8.3~\kpc$ \citep{reid14}. Unless stated otherwise, all velocities are given in the reference frame of the Galactic Centre -- from all line-of-sight velocities we subtract the Sun's radial velocity towards the Galactic Centre, which we take to be $U_\odot=14~\kms$ \citep{schoenrich12}, and from proper motions we subtract the Sun's orbital motion, which induces a proper motion of $\{\mu_l,\mu_b\}=\{-6.379,-0.202\}~{\rm mas}~\yr^{-1}$ in Galactic coordinates \citep{reid04,reid09}.

\section{Observational constraints and current orbital model}  \label{sec:obs}
In this paper, we aim to constrain the orbital motion of the gas streams seen in the CMZ. We therefore isolate only those pieces of information that are directly related to the ballistic orbital dynamics of the gas. Of course, the long-term goal is to connect the resulting picture to additional constraints and physics, such as e.g.~hydrodynamics, star formation, feedback, and the 3D geometry derived from absorption studies. Aspects of these are discussed qualitatively in \S\ref{sec:disc}.

\subsection{General properties of the dense gas in the CMZ} \label{sec:general}
\begin{figure}
\center\resizebox{\hsize}{!}{\includegraphics{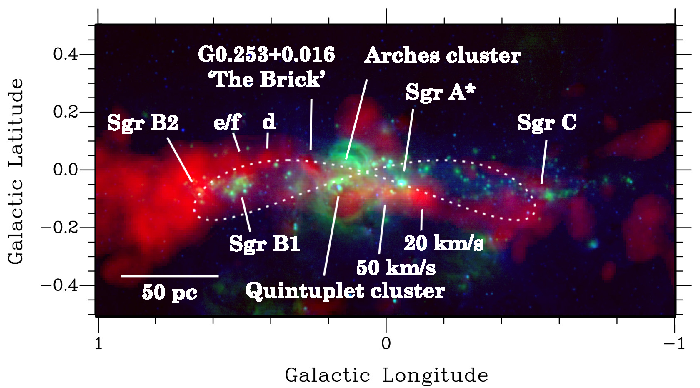}}\\
\caption[]{\label{fig:cmz}
      {Three-colour composite of the CMZ within a galactocentric radius of $R\sim150$~pc. Red shows an integrated-intensity map of the HOPS NH$_3(1,1)$ emission (see text) to indicate the gas with a density $n>{\rm several}~10^3~\cmc$, green shows the MSX 21.3$\mu{\rm m}$ image \citep{egan98,price01}, and blue shows the MSX 8.28$\mu{\rm m}$ image. The MSX data shows PAH emission (mostly tracing cloud edges), young stellar objects, and evolved stars. The dotted line shows the model of \citet{molinari11}.}
                 }
\end{figure}
Before presenting our orbital model for the dense gas in the CMZ, we first discuss the observed structure and the main existing model. The dense ($n>{\rm several}~10^3~\cmc$) gas morphology within the central degree of the CMZ (i.e.~within a galactocentric radius of $R\sim150$~pc) is shown in Figure~\ref{fig:cmz}, which in red reveals a pronounced figure-eight shape. This shape inspired the twisted-ring model of \citet{molinari11}, which is the most recent model for the structure of the CMZ (see below).

While the morphology of the molecular gas in the CMZ provides a clear picture, its observed kinematics are intricate. The gas often harbours multiple velocity components along the line of sight \citep[e.g.][]{bally88,morris96}. This complicates the dynamical analysis of the gas and obstructs a straightforward derivation of its orbital structure. In this light, a sensible starting point is a simple geometric model \citep{molinari11} that parametrizes the morphology of the gas and roughly matches its kinematics. This model is based on a constant orbital velocity and does not account for the physical dynamics of the orbital motion in the Galactic gravitational potential.

In the \citet{molinari11} model, the morphology of the gas is represented by a twisted ring, which takes the shape of an infinity symbol when projected in the plane of the sky (see Figure~\ref{fig:cmz} above, Figure~5 of \citealt{molinari11}, and Figure~5 of \citealt{kruijssen14b}). Viewed from above the Galactic plane, the model follows an ellipse, in which the bottom of the gravitational potential at Sgr~A$^*$ is positioned off-centre in the direction of the Sun. The gas orbits the ring at a constant orbital velocity of $v_{\rm orb}=80~\kms$. It approaches us head-on at Sgr~C, passes below Sgr~A$^*$, through the Brick to Sgr~B2. It then recedes from the viewer to the back side of the `ring', passing below the Brick and crossing itself in projection just below the Arches cluster, and then continues to positive Galactic latitudes, finally closing the loop at Sgr~C.

The \citet{molinari11} model has led to several important new insights. We shall see below that there are also areas where it can be improved, both in terms of its physical motivation and its agreement with the observational data. The aim of this section is to systematically identify the coherent gas structures in the observed position-velocity space. This information is used in \S\ref{sec:orbit} to construct a self-consistent orbital model of the molecular gas in the CMZ.

\subsection{Systematic survey of dense gas} \label{sec:survey}
The complex phase-space structure of the gas obstructs the straightforward identification of its orbital characteristics. We perform a systematic survey of coherent structures in position-velocity space to obtain the observational dataset that is required to fit orbital models to. We use the NH$_3(1,1)$ emission line observations from the H$_2$O southern Galactic Plane Survey \citep[HOPS,][]{walsh11,purcell12}, which traces the gas at densities $n>{\rm several}~10^3~\cmc$.

The HOPS data allows us to trace moderately high-density gas across the necessary range in Galactic longitude, from $l=-0.7^\circ$ to $0.8^\circ$, over which the gas is coherent in position-velocity space. Even though NH$_3(1,1)$ exhibits hyperfine structure, the lines are typically not detected in the CMZ clouds due to the broad linewidth of the gas. In the narrow-linewidth regions where hyperfine structure is observed, this is easily identified thanks to the known separations of these lines. The large linewidth difference between clouds in the CMZ and in the Galactic disc \citep[e.g.][]{shetty12,kruijssen13b} makes it trivial to identify contaminants. Higher-$J$ NH$_3$ transitions, such as NH$_3(3,3)$, can be brighter than NH$_3(1,1)$, but we see evidence at several locations of maser activity in the NH$_3(3,3)$ line, whereas the present work requires the kinematics of the dense thermal gas to be traced. Finally, NH$_3(1,1)$ shows no strong signs of opacity or self-absorption effects across the CMZ (except for Sgr~B2). By contrast, these effects are prevalent in lines from other molecules, such as HCO$^+$, HCN, and HNC \citep[e.g.][]{jones12}.

In summary, by using single-dish observations of a moderately high-density tracer, we reliably map the global kinematics of the molecular gas in the CMZ. Given that these bulk kinematics should be tracer-independent, we expect that the resulting orbital fit will be compatible with the observations from other molecular gas surveys, irrespective of e.g.~spatial resolution or critical density.

The position-velocity distribution of the gas is mapped as follows.
\begin{enumerate}
\item
We start at Sgr~C and follow the low-latitude stream towards increasing Galactic longitudes, which at the same time gradually extends to increasing latitudes.
\item
Every second pixel in longitude (corresponding to intervals of $\Delta l=1'$ or $\Delta x=2.5$~pc), we record the latitude $b$, line-of-sight velocity $v_{\rm los}$ and linewidth $\Delta v_{\rm los}$ of the gas stream. The uncertainties on the recorded latitudes are taken to be the resolution of the observations, i.e.~$\sigma_b\sim1'=2.5~\pc$, whereas those on the line-of-sight velocity are assumed to correspond to the velocity dispersion, i.e. $\sigma_v=\Delta v_{\rm los}/\sqrt{2\ln{2}}$. The latter uncertainties can be asymmetric around $v_{\rm los}$.
\item
At a given longitude, if there is only a single velocity component along the line of sight, we select the latitude of the pixel with the largest peak intensity. If there are multiple velocity components along the line of sight, we select the latitude of the pixel with the largest peak intensity {\it at the velocity corresponding to the coherent velocity structure of the gas stream}. This choice is made to avoid the inclusion of other gas structures along the line of sight, which in rare cases may locally outshine the emission from the gas stream, causing the pixel's peak intensity to be reached at a different velocity than the adopted one.
\item \label{pt:indep}
After the stream passes in front of Sgr~A$^*$ (at the locations of the 20 and 50~$\kms$ clouds, see Figure~\ref{fig:cmz} and \citealt{bally10}), the gas emission continues through a region with three independent velocity structures along the line of sight. Two of these connect in velocity space to the velocity of the stream, and it is not possible to establish whether this indicates a bifurcation or a chance projection. Dynamically, a bifurcation at the leading end of a gas stream is unlikely (although tidal effects could play a role), suggesting that the position-velocity structure arises from the projection of an independent component.\footnote{Another explanation for the bifurcation could be that it is driven by feedback from the Arches and Quintuplet clusters, see \S\ref{sec:gaps}.} We therefore only follow the brighter of the two branches.
\item
We continue mapping the gas across the Brick towards Sgr~B2, after which we follow the emission at low latitudes towards low longitudes, crossing the previously-mapped emission towards high latitudes, before eventually returning to the position of Sgr~C.
\end{enumerate}
Following the above procedure, we obtain the phase-space structure of the gas stream. A total of 226 data points is collected at 113 positions, providing $\{b,v_{\rm los}\}$ as a function of $l$ necessary for fitting the orbital models in \S\ref{sec:orbit}.

\begin{figure}
\center\resizebox{\hsize}{!}{\includegraphics{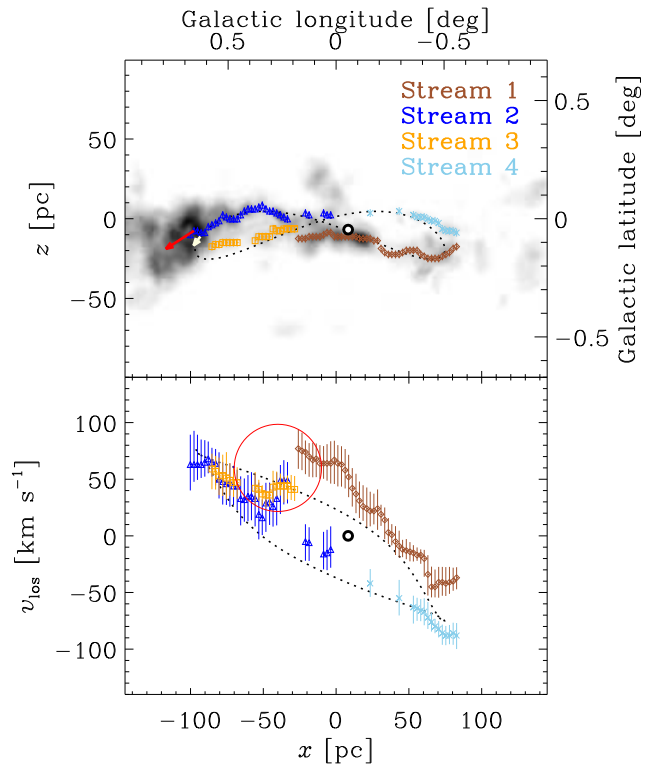}}\\
\caption[]{\label{fig:orbit_mol}
      {Comparison of the \citet{molinari11} parametric orbital model (dotted line) with the observed integrated-intensity map of NH$_3(1,1)$ emission near the Galactic Centre, tracing gas with volume densities $n>{\rm several}~10^3~\cmc$ (grey scale). Symbols with error bars show the coherent phase-space structure obtained as described in \S\ref{sec:survey}). We have divided the gas into four coherent streams in position-velocity space that are colour-coded as indicated by the legend. In the \citet{molinari11} model, the front side of the gas stream consists of Streams~1 and~2, whereas the back side consists of Streams~3 and~4. The open black circle denotes the position of Sgr~A$^*$. {\it Top panel}: Distribution in Galactic longitude and latitude $\{l,b\}$. The red arrow indicates the observed proper motion vector of Sgr~B2 and the bright yellow arrow represents the model prediction. The latter highlights the model's direction of motion, which is anti-clockwise on the left-hand side of the orbit and clockwise on the right-hand side. {\it Bottom panel}: Distribution in Galactic longitude and line-of-sight velocity $\{l,v_{\rm los}\}$. The model's direction of motion is anti-clockwise. The red circle indicates a feature in position-velocity space that is discussed in detail in \S\ref{sec:improv}.}
                 }
\end{figure}
The resulting distribution of gas in $\{l,b\}$ and $\{l,v_{\rm los}\}$ space is shown in Figure~\ref{fig:orbit_mol}, together with the parametric model of \citet{molinari11}. The data points trace the figure-eight shape that we already highlighted in Figure~\ref{fig:cmz}. The line-of-sight kinematics indicate clockwise rotation when seen from above the Galactic plane, with a clear gradient across the Galactic longitude range. Roughly speaking, the gas at positive longitudes (part of Stream~1 as well as Streams~2 and~3) is receding from our position, whereas the gas at negative longitudes (part of Streams~1 and~2 as well as Stream~4) is approaching us. The separation into these four different streams is done to ease the comparison to the orbital model in \S\ref{sec:orbit} and does not necessarily have a physical meaning.

Comparing these results to previous studies of the same region in which coherent gas streams were identified \citep[e.g.][]{bally88,binney91,sofue95,stark04}, we see that we have obtained a clean sample. For instance, \citet{sofue95} find two main `arms', the first of which is constituted by our Streams 2--4, whereas the second one represents our Stream~1. In addition, they distinguish two more arms that are offset to slightly higher longitudes (and in one case latitudes) from Sgr~A$^*$. These were already mentioned when reporting the three or four independent velocity structures along the line of sight (see point~\ref{pt:indep} above). The results of \citet{sofue95} support our decision to omit these structures due to being contaminants along the line of sight (although their large line widths do suggest that they are physically part of the CMZ). Previous studies did not identify the line-of-sight velocity discontinuity that is highlighted by the red circle in Figure~\ref{fig:orbit_mol}, which will prove crucial in our dynamical analysis (see \S\ref{sec:improv} and \S\ref{sec:orbit}).

The short red line in Figure~\ref{fig:orbit_mol} shows the observed proper motion vector of Sgr~B2 \citep{reid09}, which is the only cloud/complex in the CMZ for which such a measurement exists. The observed proper motion is $\{\mu_l,\mu_b\}=\{2.3\pm1.0,1.7\pm1.0\}~{\rm mas}~{\rm yr}^{-1}$ and represents the mean motion of two water masers.\footnote{The small number of sources implies large uncertainties, even when assuming a reasonable velocity dispersion \citep{reid09}. The quoted uncertainties are based on a velocity dispersion of $40~\kms$, but the true uncertainty may be larger if the masers trace rapid outflows or are driven by runaway stars.} The proper motion can be combined with the observed radial velocity of $v_{\rm los}=63\pm25~\kms$ to obtain a 3D orbital velocity of Sgr~B2 of $v_{\rm orb}=129\pm36~\kms$.

\subsection{Main points of improvement for a new model} \label{sec:improv}
We now discuss the comparison between the observed gas structures and the \citet{molinari11} model, with the aim of identifying the key areas in which the model can be improved upon.

Given the observed distribution of the data, we can calculate the goodness-of-fit statistic $\chi_{\rm red}^2$ for the \citet{molinari11} model. Fixing the Galactic longitude, their model has seven independent parameters (two additional parameters are fixed {\it a priori}) and is compared to 224 data points (combining the Galactic latitude and line-of-sight velocity measurements at 112 longitudes). This dataset provides an important step forward compared to that used in \citet{molinari11}, where the orbit was fitted to the position-velocity structure at 20 longitudes along the orbit (see below). In calculating the $\chi_{\rm red}^2$ statistic, we compare Streams~1 and~2 to the front side of their model (going from low longitude and latitude at Sgr~C through the Brick to Sgr~B2). Streams~3 and~4 are compared to the back side of their model (going from Sgr~B2, passing below the Brick and above Sgr~A$^*$ back to Sgr~C). We omit the proper motion of Sgr~B2 and find $\chi_{\rm red}^2=5.3$.

Considering that the \citet{molinari11} model was not originally fitted to these exact observations (\citealt{molinari11} used {\it Herschel} data in combination with CS$(1,0)$ observations by \citealt{tsuboi99} to add in the velocity information at 20 positions), the above $\chi_{\rm red}^2$ indicates a reasonable fit. Indeed, the model has two key properties that must also be present in future models, because they are essential for reproducing the observed position-velocity structure of the molecular gas in the CMZ.
\begin{enumerate}
\item
{\it The orbit is eccentric}. Two simple properties of the observed position-velocity structure show that this is required. Firstly, the line-of-sight velocities near $l=0^\circ$ and the position of Sgr~A$^*$ are non-zero. If the orbit were circular, the velocity component along the line of sight should vanish at positions that in projection are near the bottom of the gravitational potential. Secondly, the 3D space velocity of Sgr~B2 is roughly $v_{\rm orb}=129~\kms$. Given current measurements of the gravitational potential \citep[see \S\ref{sec:orbit} below]{launhardt02}, the circular velocity at the position of Sgr~B2 is $v_{\rm circ}>165~\kms$.\footnote{This is a lower limit because the separation along the line of sight between Sgr~B2 and Sgr~A$^*$ cannot be measured directly and has to be inferred from orbital modelling (see \S\ref{sec:orbit} and e.g.~\citealt{sawada04}) or X-ray absorption \citep{ryu09}. The lower limit given minimizes the galactocentric radius of Sgr~B2 by assuming that it resides at the same distance as Sgr~A$^*$.} The fact that the 3D velocity is lower at the $1\sigma$--$2\sigma$ level shows (1) that the orbit must be eccentric and (2) that Sgr~B2 resides closer to apocentre than it does to pericentre. Both of these conclusions are in accordance with the \citet{molinari11} model.
\item
{\it The orbit oscillates vertically}. Figure~\ref{fig:orbit_mol} clearly shows that some degree of vertical motion must be present -- the model reproduces the required amplitude, albeit with line-of-sight velocities that are inconsistent at the $\sim2\sigma$ level. In the \citet{molinari11} model, the ratios between the radial, azimuthal and vertical oscillation periods are $P_R:P_\phi:P_z=1:2:1$, respectively, i.e.~the orbit is closed and each orbital revolution holds two radial and vertical oscillations. Within the framework of this parametric model, it is not possible to establish whether this is the true $P_R:P_\phi:P_z$ ratio -- different combinations of the radial extent of the gas (the structure may extend beyond the positions of Sgr~B2 and Sgr~C) and the vertical flattening of the gravitational potential can give rise to similar structure between Sgr~B2 and Sgr~C (see \S\ref{sec:orbit}).
\end{enumerate}

In addition to these successes, a more detailed comparison of the \citet{molinari11} model and the NH$_3(1,1)$ observations also reveals several areas of improvement for new models (see \S\ref{sec:orbit}).
\begin{enumerate}
\item
The first of two observational questions is the origin of the discontinuity in $\{l,v_{\rm los}\}$ space indicated by the red circle in the bottom panel of Figure~\ref{fig:orbit_mol}. If a structure is coherent in position-velocity space, then the change of the line-of-sight velocity between the tangent points of the projected orbit {\it must} be monotonic. Figure~\ref{fig:orbit_mol} shows that going from Sgr~C to Sgr~B2, the velocity first increases, then decreases, before it increases again. The change occurs in the area between Sgr~A$^*$ and the Brick, which has multiple velocity components along the line of sight (see \S\ref{sec:general}--\ref{sec:survey}). However, none of these components has the appropriate velocity to fill the observed gap in $\{l,v_{\rm los}\}$ space. The inescapable conclusion is that {\it Streams~1 and~2 are not connected}.\footnote{In principle, the same could be said about Stream~2 itself, which shows an opposite velocity gradient at the low-longitude end. However, this component corresponds to the Brick, which is clearly a coherent gas structure. Because it is a single cloud, we suspect the opposite velocity gradient to be a tidal effect. In Paper~II, we will explain this feature in detail.}
\item
The second observational issue is the proper motion of Sgr~B2. The observed proper motion is $\{\mu_l,\mu_b\}=\{2.3\pm1.0,-1.4\pm1.0\}~{\rm mas}~{\rm yr}^{-1}$ and the 3D orbital velocity of Sgr~B2 is $v_{\rm orb}=126\pm37~\kms$. However, the \citet{molinari11} model predicts $\{\mu_l,\mu_b\}_{\rm M11}=\{0.57,-0.69\}~{\rm mas}~{\rm yr}^{-1}$ and $v_{\rm orb,M11}=80~\kms$ by construction. These numbers are inconsistent with the observed values at the $1\sigma$--$3\sigma$ level. We find that while the model does reproduce the motion of Sgr~B2 along the line of sight, {\it Sgr~B2 has a much larger velocity in the plane of the sky than predicted by the model}. This discrepancy has previously led to suggestions that the orbit may extend further than in the \citet{molinari11} model \citep{kruijssen14b}, or that Sgr~B2 may be located $130\pm60~\pc$ in front of Sgr~A$^*$ (\citealt{reid09}; J.~Bally, private communication).\footnote{Note that while \citet{reid09} do determine the distance to Sgr~B2 through trigonometric parallax measurements, the uncertainties of that measurement are too large ($\sim0.6~\kpc$) to establish its line-of-sight position relative to Sgr~A$^*$. The quoted $100$-$\pc$ offset ``rests on the assumption of a low eccentricity Galactic orbit for Sgr~B2'' \citep{reid09}, i.e.~on the assumption that the orbital motion is close to the circular velocity at the galactocentric radius of Sgr~B2. In \S\ref{sec:orbit}, we will show that this assumption of a near-circular orbit does not hold.}
\item
The first of three physical difficulties for the \citet{molinari11} model is that the orbit is closed. Because the mass distribution in the CMZ is extended, closed orbits are only possible if the potential is not axisymmetric. While the Galactic bar causes strong deviations from axisymmetry on $\sim\kpc$ scales, there is no evidence that such asymmetries persist down to scales as small as the 100-pc gas streams that we consider here \citep{rodriguezfernandez08}. Hence, {\it the gas likely follows an open orbit}. If the orbit is also open in the rotating reference frame of the bar, this introduces the possibility that the gas streams cross each other and interact. As we will see in \S\ref{sec:orbit}, this is unlikely to occur due to the orbit's vertical motion.
\item
The second physical problem is the assumption of a constant orbital velocity $v_{\rm orb}=80~\kms$. As stated previously, the orbit must be eccentric, which leads to a (possibly substantial) variation of the orbital velocity with the orbital phase angle. Sgr~B2 resides near apocentre in the \citet{molinari11} model, where the orbital velocity should reach its minimum. However, Sgr~B2 has a 3D space velocity of $v_{\rm orb}=126~\kms$ and the local circular velocity is even higher at $v_{\rm circ}>165~\kms$ \citep{launhardt02}. It therefore seems inevitable that the mean orbital velocity is well in excess of $80~\kms$. We conclude that {\it the orbital velocity must vary along the orbit and is likely much higher than $v_{\rm orb}=80~\kms$}.
\item
Finally, the third physical issue is that Sgr~A$^*$ does not reside at the focus of the ellipse in the \citet{molinari11} model. Even if there is a precession of the phase angles at which pericentre and apocentre occur (as is appropriate for an near-axisymmetric, extended mass distribution), {\it the orbit's focus should always coincide with the bottom of the gravitational potential}. The main argument for the skewed position of Sgr~A$^*$ in the \citet{molinari11} model is twofold. Firstly, the $50$~and $20~\kms$ clouds (the part of Stream~1 with positive line-of-sight velocities) have been suggested to be physically interacting with Sgr~A$^*$ \citep{herrnstein05}. In addition, the line-of-sight velocity difference between the $50$~and $20~\kms$ clouds could not be explained by the \citet{molinari11} model, prompting the suggestion that it may be caused by a proximity of Sgr~A$^*$ to the front side of the ellipse. However, as we have just seen, these two clouds cannot be part of the same structure as Stream~2, which greatly expands the range of their possible orbital parameters -- as we will show in the next section, self-consistent orbital solutions can be obtained in which Sgr~A$^*$ does reside at the orbit's focus.
\end{enumerate}

\section{Orbital modelling} \label{sec:orbit}
\subsection{Model setup} \label{sec:setup}
\begin{figure}
\center\resizebox{\hsize}{!}{\includegraphics{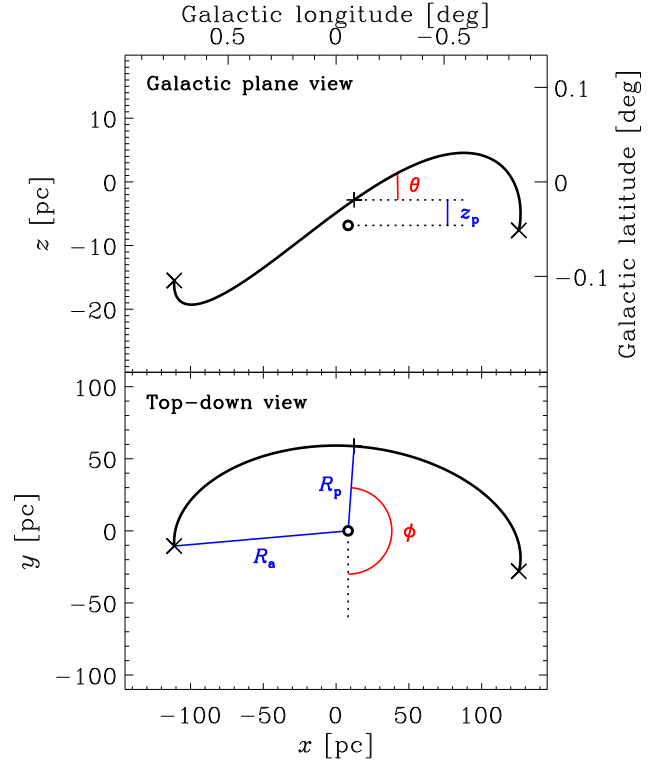}}\\
\caption[]{\label{fig:orbit_geo}
      {Two-dimensional projections of an orbital segment illustrating the free parameters used to define each orbital model. The thick black line indicates an orbital segment between two successive apocentres (crosses), passing through a single pericentre (plus symbol) in between. From the top-down perspective, the motion along this segment is in the clockwise direction, unless $\theta>\pi/2$. The open black circle denotes the position of Sgr~A$^*$. Blue lines and labels indicate distances, whereas red lines and labels represent angles. {\it Top panel}: Configuration as seen from Earth, corresponding to the Galactic longitude-latitude plane. {\it Bottom panel}: Configuration as seen from above the Galactic plane, corresponding to the Galactic longitude-line of sight plane.}
                 }
\end{figure}
We now turn to the orbital modelling of the gas structure in the CMZ. This first requires adopting a gravitational potential and an informed choice of priors for the orbital parameters. We adopt a flattened version of the potential implied by the mass distribution from \citet{launhardt02}, where the amount of flattening is left as a free parameter. The potential is described in detail in Appendix \ref{sec:potential}.

We characterise the orbits in this potential using six parameters. These are then varied to obtain a fit to the observed position-velocity data. The six parameters are as follows (see Figure~\ref{fig:orbit_geo} for a visual representation).
\begin{enumerate}
\item
The apocentre radius $R_{\rm a}$, which is varied between $R_{\rm a}=100~\pc$ (the projected separation between Sgr~B2 and Sgr~A$^*$) and $R_{\rm a}=200~\pc$. This way, we include all solutions that extend at least as far as Sgr~B2, whereas the widest orbits reach the cloud complex at $l=1.3^\circ$.
\item
The pericentre radius $R_{\rm p}$, which is varied between $R_{\rm p}=\max{(z_{\rm p})}$ (see below) and $R_{\rm p}=\min{(R_{\rm a})}$. This is the maximum allowed range based on the observed gas distribution -- the pericentre radius cannot be smaller than the vertical separation at pericentre, nor can it be larger than the apocentre radius. Together with the apocentre radius, the pericentre radius sets the total orbital velocity at the extreme ends of an eccentric orbit in a predefined potential.
\item
The height $z_{\rm p}$ above the Galactic plane at which pericentre is reached. This value is varied between $z_{\rm p}=-15~\pc$ and $z_{\rm p}=15~\pc$, which spans the projected minimum and maximum vertical separation between the gas and Sgr~A$^*$.
\item
The velocity angle at pericentre $\theta$, which indicates the angle between the velocity vector (i.e.~the orbit) and the Galactic plane during pericentre passage. This parameter is varied between $\theta=-15^\circ$ and $\theta=15^\circ$. This range of angles is adequate, because the range of possible galactocentric radii and latitudes implies that the maximum angle at any point along the orbit is $\theta_{\rm max}=\arctan{(z_{\rm max}/R_{\rm a,min})}\sim10^\circ$.
\item
The projection angle $\phi$, which reflects the angle between the vectors origin--observer and origin--pericentre, where positive values indicate a transformation in the anti-clockwise direction. We consider a range of $60^\circ$ around a prior chosen below based on the observed position-velocity structure of the streams.
\item
The vertical-to-planar axis ratio of the potential $q_{\Phi}$, which indicates the factor by which the gravitational potential is compressed in the vertical direction. Geometrically, the observed flattening seems to be well-characterised by $q_{\Phi}\sim0.5$ \citep{rodriguezfernandez08,molinari11} and we consider values between $q_{\Phi}\sim0.4$ and $q_{\Phi}\sim0.8$. This range extends from the maximum allowed flattening at low $q_\Phi$ needed to avoid negative densities (see Appendix~\ref{sec:potential}) to a near-absence of flattening at high $q_\Phi$.
\end{enumerate}
Each combination of the above six parameters defines an orbit, which we obtain by initialising it at pericentre and performing a leapfrog integration in the positive and negative time directions. Using a timestep of $\Delta t=10^3~\yr$ gives well-converged results.

The key remaining question is which parts of the orbit should be fitted to the different coherent streams identified in Figure~\ref{fig:orbit_mol}. Based on the enhanced $70\mu{\rm m}$ absorption seen in Streams~1 and~2, \citet{molinari11} conclude that these streams must be in front of the bulk of the warm dust emission. Therefore, Streams~1 and~2 constitute the front part of the gas distribution, whereas Streams~3 and~4 reside at the far side of the Galactic Centre. However, we discussed in \S\ref{sec:improv} that Streams~1 and~2 cannot be connected due to a discontinuity in their line-of-sight velocities. The velocities of Streams~3 and~4 are consistent with constituting a single structure, and Stream~2 connects smoothly to Stream~3, whereas Stream~1 connects smoothly to Stream~4. We therefore let the centre of the orbit coincide with a pericentre passage along Streams~3 and~4 (i.e.~we fit $\phi$ in the range $\phi=120$--$240^\circ$), with Stream~1 representing a downstream `tail' and Stream~2 lying upstream. The resulting order of the gas streams to which the orbits are fitted is 2-3-4-1. Finally, the direction of motion is constrained by the slope of the streams in the bottom panel of Figure~\ref{fig:orbit_mol}, which indicate rotation in the clockwise direction when observed from above the Galactic plane.

\subsection{Orbital fit and comparison to observations} \label{sec:fit}
\begin{table}
 \centering
  \begin{minipage}{32mm}
  \caption{Orbital parameters}\label{tab:fit}
  \begin{tabular}{@{}l c@{}}
  \hline 
  Parameter & Value\\
  \hline
  $R_{\rm a}$ & $121^{+15}_{-16}~\pc$ \\
  $R_{\rm p}$ & $59^{+22}_{-19}~\pc$ \\
  $z_{\rm p}$ & $4^{+6}_{-6}~\pc$ \\
  $\theta$ & $9^{+3}_{-3}~{\rm deg}$ \\
  $\phi$ & $176^{+7}_{-9}~{\rm deg}$ \\
  $q_\Phi$ & $0.63^{+0.07}_{-0.06}$ \\
  \\
  $e$ & $0.34^{+0.16}_{-0.20}$ \\
  $v_{\rm orb,a}$ & $101^{+54}_{-29}~\kms$ \\
  $v_{\rm orb,p}$ & $207^{+17}_{-20}~\kms$ \\
  $P_R$ & $2.03^{+0.70}_{-0.18}~\myr$ \\
  $P_\phi$ & $3.69^{+0.68}_{-0.30}~\myr$ \\
  $P_z$ & $2.27^{+0.70}_{-0.34}~\myr$ \\
  \hline
\end{tabular}\\
\end{minipage}
\end{table}
By varying the six orbital parameters listed above and fitting the resulting orbital models to the data obtained in \S\ref{sec:survey}, we obtain a best-fiting orbit with $\chi_{\rm red}^2=2.0$. Note that we do not include the proper motion of Sgr~B2 in the fitting process. The best-fitting parameters of our model orbit are provided in Table~\ref{tab:fit}, together with six derived properties of the orbit. These are its eccentricity $e$, the orbital velocity at apocentre $v_{\rm orb,a}$, the orbital velocity at pericentre $v_{\rm orb,p}$, the radial oscillation period $P_R$, the azimuthal oscillation period $P_\phi$, and the vertical oscillation period $P_z$. The numbers of decimals reflect the accuracy attained by the $\chi_{\rm red}^2$ minimisation. The error margins listed for the derived quantities reflect the extremes reached in the part of parameter space where $\min{(\chi_{\rm red}^2)}<\chi_{\rm red}^2<\min{(\chi_{\rm red}^2)}+1$. The dependence of the best-fitting orbit on each of the six free parameters is discussed in Appendix~\ref{sec:appendix}, and the complete orbital solution is tabulated in Appendix~\ref{sec:fullorbit}.

The orbital fitting process covers a large parameter space. Starting from the parameter ranges listed in \S\ref{sec:setup}, we iteratively narrow these ranges until the best-fitting parameter set can be identified with the desired accuracy. This way, more than $10^5$ different orbital solutions are integrated. It may be possible that a better fit can be achieved than our best-fitting orbit, but such a hypothetical solution must exist outside of the parameter range considered in the fitting process. This in itself is problematic -- in \S\ref{sec:setup}, we discuss several reasons why orbits outside the considered parameter ranges are highly unlikely or even unphysical.

As an example of a possible consideration that would have been necessary had the fitting been done by hand, we note that more extended (higher $R_{\rm a}$) orbits could be possible if the orbit extends to higher longitudes beyond the position of Sgr~B2. However, the observed vertical oscillations are only reproduced if their number per azimuthal period increases accordingly. As such, high-$R_{\rm a}$ orbits require a stronger flattening of the gravitational potential, which beyond the range already considered in the fitting process leads to unphysical solutions (see \S\ref{sec:geo} and Appendix~\ref{sec:potential}). This is exactly the type of consideration that is automatically taken care of by running a $\chi_{\rm red}^2$ minimisation.

\begin{figure}
\center\resizebox{\hsize}{!}{\includegraphics{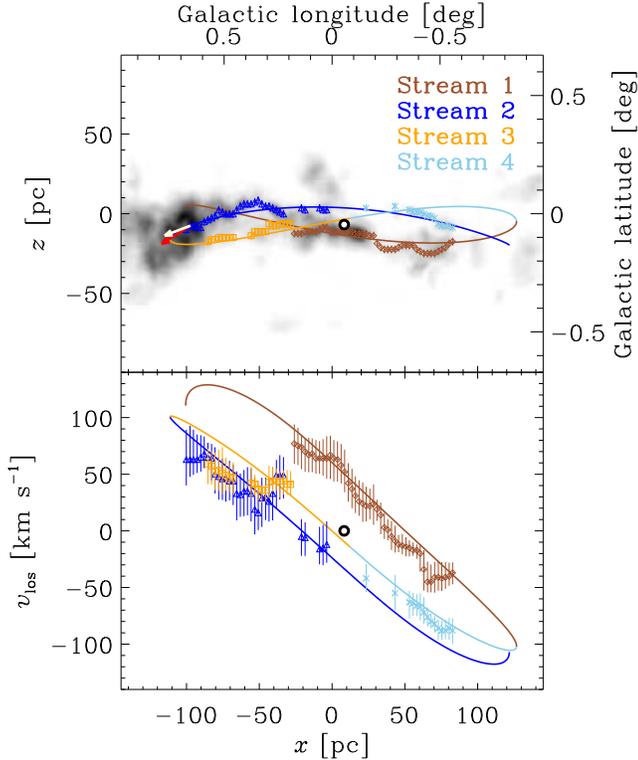}}\\
\caption[]{\label{fig:orbit_fit}
      {Comparison of our orbital model (solid line) with the observed integrated-intensity map of NH$_3(1,1)$ emission near the Galactic Centre, tracing gas with volume densities $n>{\rm several}~10^3~\cmc$ (grey scale). Symbols with error bars show the coherent phase-space structure (see \S\ref{sec:survey}) to which the model was fitted. We have divided the gas into four coherent streams in position-velocity space that are colour-coded as indicated by the legend. In our model, the back side of the gas stream consists of Streams~3 and~4, whereas Streams~1 and~2 represent the two (independent) ends of the stream on the front side. The open black circle denotes the position of Sgr~A$^*$. The model starts at Stream~2 (the overlap with Stream~4 is coincidental but could fit too) and continues through Streams~3 and~4 to Stream~1. {\it Top panel}: Distribution in Galactic longitude and latitude $\{l,b\}$. The red arrow indicates the observed proper motion vector of Sgr~B2 and the bright yellow arrow represents the model prediction. {\it Bottom panel}: Distribution in Galactic longitude and line-of-sight velocity $\{l,v_{\rm los}\}$.}
                 }
\end{figure}
\begin{figure}
\center\resizebox{\hsize}{!}{\includegraphics{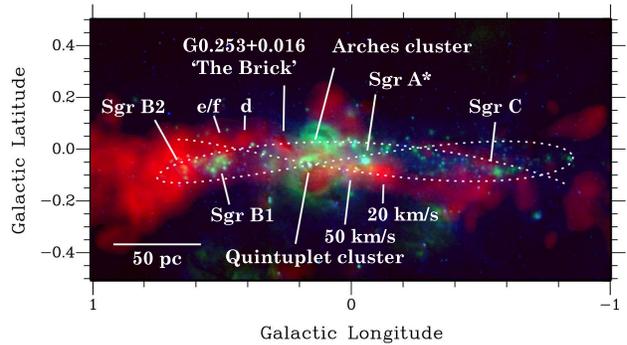}}\\
\caption[]{\label{fig:cmzfit}
      {Repeat of the three-colour composite of the CMZ from Figure~\ref{fig:cmz}. This time, the white dotted line shows our best-fitting orbit. Note that the clouds near $l=-1^\circ$ have $v_{\rm los}\sim\{0,140\}~\kms$ (bottom and top, respectively) and hence are not associated with the gas stream that the model was fitted to.}
                 }
\end{figure}

The best-fitting orbit is compared to the observations in Figure~\ref{fig:orbit_fit}. In addition, Figure~\ref{fig:cmzfit} shows the orbit overlaid on the three-colour composite image of Figure~\ref{fig:cmz}. The orbit successfully reproduces several key properties of the observed gas distribution.
\begin{enumerate}
\item
All four identified gas streams are described with a single orbit, which is consistent with the observations at the $<2\sigma$ level for most points along the orbit. The only exception is the position of Stream~4, but given the systematic uncertainties involved in the 3D shape of the gravitational potential (e.g.~the ill-constrained vertical shape and possible deviations from axisymmetry, see \S\ref{sec:questions}), the match is remarkably good across $\{l,b,v_{\rm los}\}$ space, as indicated by $\chi_{\rm red}^2=2.0$.
\item
The line-of-sight velocities near Sgr~A$^*$ are non-zero, signifying an eccentric orbit.
\item
The orbit oscillates vertically with a period close to half the azimuthal oscillation period, i.e.~$P_z/P_\phi=0.6$. Note that the radial oscillation period $P_R$ is similar to $P_z$.
\item
The discontinuity in the $\{l,v_{\rm los}\}$ plane that is indicated with the red circle in Figure~\ref{fig:orbit_mol} is accounted for by modelling Streams~1 and~2 as opposite, unconnected tails of a single, long gas stream, which wraps around the Galactic Centre in between Streams~1 and~2 (through Streams~3 and~4).
\item
The modelled proper motion of Sgr~B2 is $\{\mu_l,\mu_b\}_{\rm pred}=\{2.14,-0.75\}~{\rm mas}~{\rm yr}^{-1}$, which with $v_{\rm los,pred}=83~\kms$ gives $v_{\rm orb,pred}=124~\kms$. Comparing to the observed values of $\{\mu_l,\mu_b\}=\{2.3\pm1.0,-1.4\pm1.0\}~{\rm mas}~{\rm yr}^{-1}$, $v_{\rm los}=63\pm25~\kms$, and $v_{\rm orb}=126\pm37~\kms$, we see that all predicted velocities agree with the observed motion of Sgr~B2 at the $\la1\sigma$ level. Our orbital fit thus confirms a high orbital velocity at the position of Sgr~B2 ($v_{\rm orb,pred}=124~\kms$) and hence the proper motion of Sgr~B2 {\it does not require an orbit that extends much further than its present Galactic longitude} (which was thought previously due to the low orbital velocity of the \citealt{molinari11} model), {\it nor does it require Sgr~B2 to be situated much closer to the observer than Sgr~A$^*$} (\citealt{reid09}; J.~Bally, private communication).
\item
As mentioned in \S\ref{sec:survey}, the region that lies in projection between the Brick and the Quintuplet cluster contains three independent velocity structures along the line of sight, which are associated in position-velocity space. Two of these are explained by our model -- they represent the front and back sides of the streams (i.e.~Stream~2 and Stream~3, respectively), which are connected in position-velocity space but separated by more than $100~\pc$ along the line of sight (see \S\ref{sec:clouds} below). The third component connects to the gas that can be seen at high latitudes near $l=0^\circ$ in Figures~\ref{fig:cmz}--\ref{fig:cmzfit} and is likely physically unrelated to the streams under consideration here, unless it is being ejected from one of the streams by feedback. This possibility is underlined by its similar line-of-sight velocity to the Arches and Quintuplet clusters, implying that it could originally have been associated with the gas that is currently occupying Stream~1 (see \S\ref{sec:ymc}). We also note that this third component is physically associated with the infrared shells blown by the Arches cluster (visible in green in Figures~\ref{fig:cmz} and~\ref{fig:cmzfit}). The complex position-velocity structure surrounding the Brick has previously been proposed to result from a cloud-cloud collision \citep{lis98,johnston14}, but the results of our model show that such an event is not necessary to explain the observed gas kinematics -- instead, they are caused by the line-of-sight projection of three unrelated components.\footnote{A peak of shock tracer emission near the low-$\{l,b\}$ side of the Brick has been put forward to support the idea that the Brick has undergone a cloud-cloud collision \citep{johnston14}. However, the increased sensitivity of ALMA shows that there is no single locus of enhanced shock tracer emission \citep{rathborne14c} -- the cloud is so turbulent that it is brightly emitting throughout. Such widespread emission should follow naturally from local gravitational collapse or a compression caused by a recent pericentre passage (see \S\ref{sec:clouds}). In addition, some physical interaction between components could be possible due to the tidal stripping expected to occur at pericentre or the clearing of molecular gas shells by feedback from the Arches and Quintuplet clusters (see \S\ref{sec:gaps}). These interactions are much less dramatic than the previously-proposed collisions between gas streams or clouds, but they could still contribute to the shock tracer emission in the region.}
\end{enumerate}

In addition to the above observational points, our orbital model is dynamical rather than parametric. It therefore also satisfies several physical requirements that were unaccounted for in previous models.
\begin{enumerate}
\item
The extended mass distribution in the CMZ results in an orbit that is open rather than closed. The dissimilarity of the radial, azimuthal and vertical oscillation periods implies that the gas structure can survive on this orbit for multiple revolutions without being disrupted by self-interaction (see \S\ref{sec:clouds}).
\item
The non-zero eccentricity results in a variable orbital velocity, ranging from $v_{\rm orb}\sim100~\kms$ at apocentre to $v_{\rm orb}\sim200~\kms$ at pericentre. This is substantially higher than previous estimates. As a result, the three orbital periods are a factor of $1.4$--$1.8$ shorter than in the model of \citet{molinari11}, who obtained $\{P_{R,{\rm mol}},P_{\phi,{\rm mol}},P_{z,{\rm mol}}\}=\{3.2,6.4,3.2\}~\myr$.\footnote{Note that these periods differ from those quoted in \S3.3 of \citet{molinari11}, which actually correspond to the semi-periods of their model.}
\item
The bottom of the gravitational potential (assumed to lie at the position Sgr~A$^*$) coincides with the orbit's focus. The previous argument to move Sgr~A$^*$ towards the front side of the orbit was the fact that the line-of-sight velocity difference between the $50$~and $20~\kms$ clouds could not be explained by the \citet{molinari11} model. In our new model, the variable orbital velocity naturally leads to the observed velocity difference. It is therefore no longer necessary to displace Sgr~A$^*$.
\end{enumerate}

Other, more fundamental open questions include the existence of gaps in the observed gas distribution in comparison to our orbital model, as well as the validity of the adopted gravitational potential. These points are considered in \S\ref{sec:questions}.

\section{Implications for CMZ clouds} \label{sec:impl}
\begin{table*}
 \centering
  \begin{minipage}{127mm}
  \caption{Galactocentric radii, orbital velocities, and times relative to pericentre, apocentre, and the Brick}\label{tab:t}
  \begin{tabular}{@{}l c c c c c c c@{}}
  \hline 
  Object & $R$ & $v_{\rm orb}$ & $\Delta t_{\rm p,last}$ & $\Delta t_{\rm p,next}$ & $\Delta t_{\rm a,last}$ & $\Delta t_{\rm a,next}$ & $\Delta t_{\rm Brick}$ \\
  \hline
  Brick      & $77^{+50}_{-14}$ & $183^{+15}_{-20}$ & $0.30^{+0.30}_{-0.03}$ & $-1.73^{+0.16}_{-0.58}$ & $1.31^{+0.66}_{-0.10}$ & $-0.72^{+0.10}_{-0.22}$ & $0.00$ \\
  cloud~d    & $90^{+43}_{-11}$ & $164^{+24}_{-28}$ & $0.45^{+0.29}_{-0.05}$ & $-1.58^{+0.18}_{-0.59}$ & $1.46^{+0.64}_{-0.08}$ & $-0.58^{+0.11}_{-0.24}$ & $0.14^{+0.02}_{-0.02}$ \\
  cloud~e    & $96^{+41}_{-10}$ & $155^{+30}_{-33}$ & $0.51^{+0.29}_{-0.05}$ & $-1.52^{+0.19}_{-0.61}$ & $1.52^{+0.64}_{-0.06}$ & $-0.51^{+0.13}_{-0.26}$ & $0.21^{+0.03}_{-0.03}$ \\
  cloud~f    & $97^{+40}_{-9}$ & $152^{+31}_{-36}$ & $0.53^{+0.29}_{-0.05}$ & $-1.50^{+0.21}_{-0.61}$ & $1.54^{+0.64}_{-0.06}$ & $-0.50^{+0.13}_{-0.26}$ & $0.22^{+0.05}_{-0.03}$ \\
  Sgr~B2     & $112^{+32}_{-9}$ & $124^{+48}_{-47}$ & $0.74^{+0.26}_{-0.06}$ & $-1.30^{+0.37}_{-0.66}$ & $1.74^{+0.59}_{-1.73}$ & $-0.29^{+0.24}_{-1.63}$ & $0.43^{+0.22}_{-0.08}$ \\
  Sgr~C      & $94^{+53}_{-12}$ & $157^{+12}_{-28}$ & $1.55^{+0.37}_{-0.22}$ & $-0.48^{+0.05}_{-0.58}$ & $0.53^{+0.10}_{-0.22}$ & $-1.50^{+0.11}_{-0.93}$ & $3.28^{+0.77}_{-0.34}$ \\
  $20~\kms$  & $67^{+67}_{-20}$ & $197^{+17}_{-23}$ & $1.86^{+0.38}_{-0.14}$ & $-0.18^{+0.05}_{-0.53}$ & $0.83^{+0.10}_{-0.18}$ & $-1.20^{+0.16}_{-0.88}$ & $3.58^{+0.80}_{-0.29}$ \\
  $50~\kms$  & $62^{+67}_{-20}$ & $204^{+16}_{-22}$ & $1.94^{+0.38}_{-0.14}$ & $-0.10^{+0.05}_{-0.53}$ & $0.91^{+0.08}_{-0.18}$ & $-1.12^{+0.16}_{-0.88}$ & $3.66^{+0.80}_{-0.29}$ \\
  \hline
\end{tabular}\\
Radii are listed in $\pc$, velocities in $\kms$, and times in $\myr$.
\end{minipage}
\end{table*}
\subsection{Cloud evolution and the physics of star formation} \label{sec:clouds}
\begin{figure}
\center\resizebox{\hsize}{!}{\includegraphics{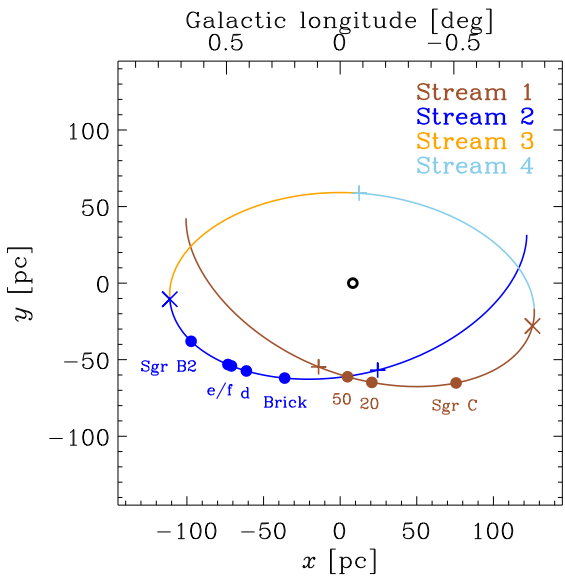}}\\
\caption[]{\label{fig:orbit_td}
      {Top-down view of our orbital model (solid line), with the observer located in the negative-$y$ direction. As in Figures~\ref{fig:orbit_mol} and~\ref{fig:orbit_fit}, the colours refer to the four coherent streams in position-velocity space. The dots indicate the implied positions in the Galactic plane of several GMCs and cloud complexes in the CMZ, the plus symbols indicate pericentres, the crosses mark apocentres, and the open black circle denotes the position of Sgr~A$^*$. Note that we only show the part of the orbit that is currently associated with the observed gas structure -- in the future, the gas stream (or its star formation products) will continue on the same rosetta-like orbit, beyond the end of Stream~1 shown here.}
                 }
\end{figure}
The orbital solution of \S\ref{sec:fit} allows us to consider the GMCs and cloud complexes in the CMZ in the context of their dynamical history. Figure~\ref{fig:orbit_td} shows a top-down perspective of the orbit, along with the implied positions in the Galactic plane of the main GMCs in the CMZ. One of the robust conclusions of our orbital parameter survey is that the `dust ridge' sequence of GMCs between the Brick and Sgr~B2 recently underwent a pericentre passage \citep[as required in the scenario of][]{longmore13b}. In addition, the $50$~and $20~\kms$ clouds are the closest to Sgr~A$^*$ of all objects under consideration here, but not quite as close as they were in previous models \citep[$R<20~\pc$,][]{molinari11}.\footnote{Another difference with respect to previous work is the orientation of the orbit. In the papers by \citet{molinari11} and \citet{johnston14}, a closed, elliptical orbit is rotated around the orbital rotation axis such that it is tilted with respect to the line of sight (i.e.~the apocentres occur away from $y=0~\pc$). As is shown in Figure~\ref{fig:orbit_td}, our orbital model exhibits little such rotation, with both apocentres close to $y=0~\pc$. This difference arises because the orbital velocity varies in our model, allowing a good fit at all longitudes without the need of boosting the line-of-sight velocity by rotating the model.}

Comparing Figure~\ref{fig:orbit_td} to Figure~\ref{fig:orbit_fit}, we see that the independent streams never approach each other closely. From the top-down perspective in Figure~\ref{fig:orbit_td}, there are three crossings, of Streams $\{1,2\}$, $\{1,3\}$ and $\{2,4\}$, at longitudes of $l=\{-0.8,0,0.7\}^\circ$. After identifying these positions in the top panel of Figure~\ref{fig:orbit_fit}, it is clear that at none of these the streams are close in latitude, with distances consistently $\Delta z> 20~\pc$. This is consistent with the statistical behaviour of the orbit -- the ratios between the vertical, radial and azimuthal oscillation periods are non-integer, implying that the orbit does not regularly intersect with itself. For the same reason though, there must also be a time when streams do cross in three dimensions. However, their total length does not exceed 1--1.5 azimuthal orbits, indicating that such self-interactions must be extremely rare. Only tidally stripped or feedback-ejected material could regularly interact with other streams. For instance, the stripping of the Stream~2 GMCs during their pericentre passage may affect the gas orbiting on Stream~1.

We proposed in \citet{longmore13b} that the recent pericentre passage of the Stream~2 GMCs (i.e.~the Brick, clouds d/e/f, and Sgr~B2) caused them to be vertically compressed by tidal forces, leading to an accelerated dissipation of turbulent energy and eventually gravitational collapse.\footnote{While cloud collapse is taking place, the outer layers may be tidally stripped. This would be observable along the line of sight as the expansion of the cloud's outer layers, which likely have elevated temperatures and magnetic field strengths compared to the collapsing centre \citep{rathborne14,bally14}.} The exciting prospect of this scenario is that the GMCs on Stream~2 may follow an absolute-time sequence of contraction and star formation, in which the zero point of each cloud's evolution coincides with the moment of its pericentre passage. While the detailed physics of the tidal compression and turbulent dissipation will be considered in Paper~II using numerical simulations, we can already use our orbital model to determine the time since pericentre for each of the GMCs. This then defines the absolute timeline on which the cloud evolution should proceed if the scenario of \citet{longmore13b} holds.

Table~\ref{tab:t} lists the 3D galactocentric radii and orbital velocities of the objects in Figure~\ref{fig:orbit_td}, as well as the times since (until) their last (next) pericentre and apocentre passages  ($\Delta t_{\rm p}$ and $\Delta t_{\rm a}$, respectively). We see that the Brick experienced its last pericentre passage $\Delta t_{\rm p,last}=0.30~\myr$ ago, whereas Sgr~B2 is closer to its upcoming apocentre passage and has $\Delta t_{\rm p,last}=0.74~\myr$. The final column of Table~\ref{tab:t} lists the time separation between each cloud and the Brick. This is done because the uncertainties on the individual times since pericentre can be substantial, but the covariance of these implies that the time differences between the clouds are constrained much better. We find that the time elapsed since the position of the Brick is generally constrained to within $0.05~\myr$ for the clouds on the dust ridge (i.e.~from the Brick to Sgr~B2), and the time difference between the Brick and Sgr~B2 is $\Delta t_{\rm Brick}=0.43^{+0.22}_{-0.08}~\myr$.

The star formation activity in the two regions could not differ more -- the Brick is largely devoid of ongoing star formation \citep{kauffmann13}, whereas Sgr~B2 is one of the most actively star-forming protoclusters in the Local Group \citep{bally10}. In the context of the \citet{longmore13b} scenario, this indicates that once collapse is triggered, the evolution towards prevalent star formation proceeds rapidly in these clouds, taking about $0.5~\myr$. This is twice as fast as estimated previously using the \citet{molinari11} model \citep{longmore13b}, and corresponds to about one free-fall time -- the GMCs in Stream~2 have densities of $n\sim10^4~\cmc$ \citep{longmore13} and hence $t_{\rm ff}=0.34~\myr$. In this model, {\it the Brick and Sgr~B2 are separated by a single free-fall time of evolution}.

Clouds d/e/f are situated at locations intermediate to the Brick and Sgr~B2. Based on the presence of a methanol maser, which indicates that massive star formation is currently in progress \citep{immer12b}, the star formation activity of these GMCs is also at an intermediate level between the Brick and Sgr~B2. Given the time-scales listed in Table~\ref{tab:t}, the existence of these GMCs allows the detailed study of the star formation process at $\Delta t\sim0.1~\myr$ resolution (i.e.~a fraction of a free-fall time).

Previous work has shown that the CMZ globally forms stars at a rate below galactic star formation relations \citep{longmore13}. A combination of physical mechanisms is likely responsible -- crucially, much of the gas is not self-gravitating due to the extreme turbulent pressure \citep{kruijssen14b}. In previous work, we therefore proposed that the rate-limiting factor is the slow evolution of gas clouds {\it towards} collapse, which first requires the clouds to become self-gravitating and dissipate the turbulent energy. After collapse has been initiated, star formation should proceed at a normal (rapid) rate. The short time interval spanned by the widely different evolutionary stages of the Brick and Sgr~B2 supports this global picture of star formation in the CMZ.

\subsection{Predictions for future observational tests} \label{sec:pred}
\begin{table*}
 \centering
  \begin{minipage}{125mm}
  \caption{Predicted proper motions in different coordinate systems}\label{tab:pm}
  \begin{tabular}{@{}l c c c c c c@{}}
  \hline 
  Object & $\mu_l$ & $\mu_b$ & $\mu_l'$ & $\mu_b'$ & $\mu_x'$ & $\mu_y'$ \\
  \hline
  Brick      & $4.51^{+0.36}_{-0.55}$ & $-0.30^{+0.68}_{-0.39}$ & $-1.87^{+0.36}_{-0.55}$ & $-0.50^{+0.68}_{-0.39}$ & $-0.54^{+0.33}_{-0.84}$ & $-1.86^{+0.47}_{-0.53}$ \\
  cloud~d    & $3.87^{+0.55}_{-0.65}$ & $-0.56^{+0.65}_{-0.35}$ & $-2.51^{+0.55}_{-0.65}$ & $-0.77^{+0.65}_{-0.35}$ & $-0.65^{+0.34}_{-0.78}$ & $-2.54^{+0.70}_{-0.67}$ \\
  cloud~e    & $3.52^{+0.70}_{-0.78}$ & $-0.65^{+0.59}_{-0.32}$ & $-2.86^{+0.70}_{-0.78}$ & $-0.85^{+0.59}_{-0.32}$ & $-0.76^{+0.36}_{-0.78}$ & $-2.88^{+0.84}_{-0.78}$ \\
  cloud~f    & $3.43^{+0.72}_{-0.88}$ & $-0.66^{+0.57}_{-0.31}$ & $-2.95^{+0.72}_{-0.88}$ & $-0.86^{+0.57}_{-0.31}$ & $-0.79^{+0.36}_{-0.77}$ & $-2.97^{+0.86}_{-0.88}$ \\
  Sgr~B2     & $2.14^{+1.27}_{-2.08}$ & $-0.75^{+0.45}_{-0.21}$ & $-4.24^{+1.27}_{-2.08}$ & $-0.95^{+0.45}_{-0.21}$ & $-1.38^{+0.57}_{-1.33}$ & $-4.12^{+1.29}_{-1.75}$ \\
  Sgr~C      & $3.81^{+0.36}_{-0.82}$ & $-0.09^{+0.35}_{-0.54}$ & $-2.57^{+0.36}_{-0.82}$ & $-0.29^{+0.35}_{-0.54}$ & $-1.09^{+0.28}_{-0.41}$ & $-2.34^{+0.38}_{-0.87}$ \\
  $20~\kms$  & $4.79^{+0.31}_{-0.58}$ & $0.57^{+0.39}_{-0.50}$ & $-1.59^{+0.31}_{-0.58}$ & $0.37^{+0.39}_{-0.50}$ & $-1.15^{+0.39}_{-0.39}$ & $-1.17^{+0.34}_{-0.65}$ \\
  $50~\kms$  & $4.82^{+0.30}_{-0.59}$ & $0.71^{+0.33}_{-0.54}$ & $-1.56^{+0.30}_{-0.59}$ & $0.51^{+0.33}_{-0.54}$ & $-1.25^{+0.42}_{-0.35}$ & $-1.06^{+0.31}_{-0.69}$ \\
  \hline
\end{tabular}\\
Proper motions are listed in ${\rm mas}~{\rm yr}^{-1}$.
\end{minipage}
\end{table*}
Next to the obvious comparison of our model with the position-velocity structure and star formation activity of well-studied GMCs, the model can also be used to make a number of predictions that can be tested in future observational comparisons.
\begin{enumerate}
\item
Perhaps the most fundamental prediction of our model is a set of proper motions for several of the GMCs in the CMZ. As discussed in \S\ref{sec:fit}, the only observational proper motion measurement presently available is that of Sgr~B2. We list our predictions for Sgr~B2 and the other GMCs in Table~\ref{tab:pm}, which facilitates a direct comparison of our model to future proper motion measurements. Note that the `primed' variables are directly observable as they include the proper motion induced by the Sun's orbital motion $\{\mu_l,\mu_b\}_\odot=\{-6.379,-0.202\}~{\rm mas}~\yr^{-1}$ \citep{reid04,reid09}, for instance $\{\mu_l',\mu_b'\}\equiv\{\mu_l,\mu_b\}+\{\mu_l,\mu_b\}_\odot$. The variables $\{\mu_x',\mu_y'\}$ indicate the proper motion in the eastward and northward directions, which in units of right ascension and declination become $\mu_\alpha'=\mu_x'/\cos{\delta}$ (for the CMZ, $\delta\sim29^\circ$) and $\mu_\delta'=\mu_y'$. The proper motions in Table~\ref{tab:pm} show that Sgr~B2 has the largest observable proper motion of all clouds listed here, followed by clouds d/e/f and Sgr~C. Most observable (i.e.~primed) proper motions are smaller than those in the reference frame of the Galactic Centre, because they are largely cancelled by the proper motion induced by the Sun's orbital motion. However, the Sun moves in the opposite direction of the gas on the far side of the gas structure (referred to as Streams~3 and~4 in this paper), which should therefore have a proper motion much larger than those listed in Table~\ref{tab:pm}. Indeed, the full orbital solution provided in Appendix~\ref{sec:fullorbit} shows that proper motions over $10~{\rm mas}~\yr^{-1}$ could be detected in the gas passing behind Sgr~A$^*$. Given that the proper motion of Sgr~B2 could be measured using a $\sim1~\yr$ baseline \citep{reid09}, the proper motion of the far-side clouds should be detectable with ease, provided that suitable masers can be identified.
\item
The vertical tidal compression of gas during pericentre passage is a robust and well-known concept, which should occur at each of the pericentre passages in Figure~\ref{fig:orbit_td}. This may explain the high column densities of the Brick and the $50$~and $20~\kms$ clouds, all of which are separated from a recent or impending pericentre passage by $\Delta t\leq0.3~\myr$. Along the same lines, we predict the presence of high-column density gas in Stream~1 at the longitudes of clouds d/e/f and Sgr~B2 with a high line-of-sight velocity ($v_{\rm los}\sim120~\kms$). A quick inspection of the HOPS NH$_3(1,1)$ data shows that there are indeed indications of such an extension, at roughly the correct latitudes ($b\sim-0.05^\circ$ versus the predicted $b\sim0.02^\circ$, also see e.g.~Figure~7 of \citealt{jones12} and Figure~7 of \citealt{ott14}). It may be problematic to detect high-column density gas near the third pericentre passage, on the far side of the gas structure (where Streams~3 and~4 meet), because it lies behind Sgr~A$^*$ along the line of sight. None the less, a strong NH$_3(1,1)$ peak with the correct, low line-of-sight velocity ($v_{\rm los}\sim15\pm10~\kms$) is present at $\{l,b\}=\{0.02,-0.02\}^\circ$, which is again consistent with our orbital model.
\item
If gravitational collapse is indeed triggered by a pericentre passage, then Sgr~B2 should not be the only region of elevated star formation activity. The two other pericentre passages in Figure~\ref{fig:orbit_td} may induce additional star formation `hotspots'. In particular, one would expect ongoing star formation activity in Stream~4 at the longitude of Sgr~C (but at higher latitudes), as well as at the tip of Stream~1 near the longitude of Sgr~B2 (compare Figure~\ref{fig:orbit_fit}). At the former location, Figure~\ref{fig:cmz} does indeed show a large concentration of young stellar objects, and \citet{immer12b} identify two H{\sc ii} regions (`D' and `E') that are located in projection on top of our Stream~1, just below cloud~d at longitudes $l=0.3$--$0.4^\circ$. Likewise, the well-known H{\sc ii} region Sgr~B1 (shown in green below clouds~e/f in Figure~\ref{fig:cmzfit}) may be downstream from Sgr~B2, thus representing a more advanced evolutionary stage in the star formation process. While these observations may provide tentative support for our model, a more conclusive picture may emerge when new proper motion data becomes available.
\item
At some unspecified time after the peak star formation activity, a population of unembedded, young stars should be present. The substantial population of 24$\mu{\rm m}$ sources at the low longitudes beyond Sgr~C (to the right in Figures~\ref{fig:cmz} and~\ref{fig:orbit_fit}, also see Figure~1 of \citealt{kruijssen14b} for longitudes $l<-1^\circ$) suggests that this point may be reached as early as apocentre. If true, a similar population may exist just downstream from the position of Sgr~B2, which corresponds to the location of the other apocentre in our model. Unfortunately, the straightforward verification of this prediction is obstructed by the extreme concentration of high-density gas along the line-of-sight, suggesting that free-free emission may provide a better test than 24$\mu{\rm m}$ observations.
\end{enumerate}
The main uncertainty associated with these predictions is the interrupted nature of the gas streams -- they exhibit gaps of low emission, possibly indicating that the supply of gas through pericentre is likely not continuous (also see \S\ref{sec:gaps}). Catching a Lagrangian mass element at any of the three phases discussed above (high-column density gas, active star formation, and unembedded young stars) could therefore depend on whether or not a concentration of gas passed through pericentre at the right time to be presently observable.\footnote{Based on the time-scales listed in Table~\ref{tab:t}, the shortest of these phases is likely the actively star-forming phase, in which case highly active regions like Sgr~B2 may be rare occurrences.} In addition, the potential of a pericentre passage to trigger gravitational collapse and star formation depends on the density and velocity dispersion of the gas. Both quantities vary along the gas stream, adding another source of stochasticity. The predicted hotspots therefore represent regions of an elevated {\it probability}, integrated over several orbital revolutions, of detecting high-column density gas, active star formation, or unembedded, young stars.

\section{Discussion} \label{sec:disc}

\subsection{Summary}
We have presented a new model for the orbital dynamics of GMCs in the central $R\la100~\pc$ of the CMZ, with the aim of characterising the time-evolution of the GMCs that follow the orbit. It is the first orbital model that accounts for the appropriate gravitational dynamics, based on the most accurate gravitational mass distribution in the CMZ that is currently available \citep{launhardt02}. The main results of this work are as follows.
\begin{enumerate}
\item
The orbit is fitted to the observed NH$_3(1,1)$ emission (\S\ref{sec:obs}; tracing gas with densities $n>{\rm several}~10^3~\cmc$) by varying (1) the apocentre radius, (2) the pericentre radius, (3) the height above the Galactic plane during pericentre, (4) the angle between the orbit and the Galactic plane during pericentre, (5) the angle between the line of sight and the vector origin-pericentre, and (6) the vertical-to-planar axis ratio of the gravitational potential. The best-fitting parameters yield a satisfactory solution ($\chi_{\rm red}^2=2.0$) and they are summarized in Table~\ref{tab:fit} (\S\ref{sec:orbit}).
\item
The best-fitting orbit reproduces the key properties of the observed gas distribution (\S\ref{sec:fit}). (1) It reproduces the observed position-velocity structure of four independent gas streams with a single orbital model. (2) It is eccentric and oscillates vertically at the rate required by the observations. (3) It reproduces a discontinuity in position-velocity space that was unaccounted for in previous models. (4) It reproduces the 3D space velocity of Sgr~B2.
\item
The physical properties of our new orbital solution differ from previous models (\S\ref{sec:fit}). (1) The orbit is open rather than closed, owing to the extended mass distribution in the CMZ. (2) The orbital velocity varies in the range $v_{\rm orb}=100$--$200~\kms$, which is higher than in previous models and gives orbital periods shorter by a factor of $1.4$--$1.8$. (3) The bottom of the gravitational potential coincides with the focus of the (eccentric) orbit.
\item
We confirm the suggestion of \citet{longmore13b} that the `dust ridge' sequence of GMCs between the Brick and Sgr~B2 recently underwent a pericentre passage, which may have triggered their collapse (\S\ref{sec:clouds}). This sequence of GMCs (the Brick, clouds~d/e/f, Sgr~B2) covers a mere $\Delta t=0.43~\myr$ (Table~\ref{tab:t}). Considering that the free-fall time at these densities ($n\sim10^4~\cmc$) is $t_{\rm ff}=0.34~\myr$, our model suggests that the quiescent and massive cloud the `Brick' and the rapidly star-forming complex Sgr~B2 are {\it separated by a single free-fall time of evolution}. This lends support to the idea that while the CMZ globally forms stars at a rate below galactic star formation relations \citep{longmore13}, which is likely due to the fact that much of the gas is not self-gravitating \citep{kruijssen14b}, star formation does proceed at a normal (rapid) rate in the self-gravitating clouds where most of the star formation occurs.
\item
Using the best-fitting orbital model and assuming that the orbital position-velocity space is filled entirely, we predict in which other regions of the CMZ one should expect an elevated probability of detecting high-column density gas, ongoing star formation, and unembedded, young stars (\S\ref{sec:pred}). We also provide proper motions of the main clouds considered in this paper (Table~\ref{tab:pm}), as well as those along the complete orbital solution (Appendix~\ref{sec:fullorbit}). These predictions should enable future studies to test our model.
\end{enumerate}

\subsection{Model assumptions and open questions} \label{sec:questions}
While the model presented in this paper provides a good fit to several of the main observed features of the gas in the CMZ, it relies on a number of assumptions and leaves several open questions. In this section, we discuss the influence on our results of the adopted gravitational potential (\S\ref{sec:pot}) and geometry (\S\ref{sec:geo}), as well as the relation of our orbital model to other constraints on the geometry of the CMZ (\S\ref{sec:geolit}), the possible origin of the observed asymmetry and gaps in the gas structure (\S\ref{sec:gaps}), the physical nature of the stream(s) (\S\ref{sec:nature}), and the relation of the gas stream to the Arches and Quintuplet clusters (\S\ref{sec:ymc}).

\subsubsection{The gravitational potential} \label{sec:pot}
We have adopted a modified form of the potential implied by the mass distribution derived by \citet{launhardt02}, which does not allow deviations from axisymmetry. The original potential is spherically symmetric, which we compressed vertically to account for some (fitted) degree of flattening (see Appendix~\ref{sec:potential}). The mass distribution of \citet{launhardt02} is the most accurate one currently available for the central ${\rm few}~100~\pc$ of the Milky Way, which restricts our analysis to the use of a simplified, modified-spherically symmetric potential. This assumption is important because deviations from axisymmetry will affect the orbital structure. However, there is no direct evidence for deviations from axisymmetry at the small radii ($R\la100~\pc$) under consideration in this paper \citep{rodriguezfernandez08}.\footnote{This does not mean that the influence of larger-scale asymmetries cannot affect the inner CMZ at all -- \citet{bissantz03} show that the very inner resonant `$x_2$' orbits that are caused by the Galactic bar may have pericentre radii as low as $R_{\rm p}\sim20~\pc$.} Unfortunately, a more conclusive picture will require the detection of azimuthal variations in the CMZ's gravitational potential, which is hard to achieve. Accurate proper motion measurements with Gaia and ALMA may help to resolve this issue.

We have made the additional assumption that Sgr~A$^*$ coincides with the bottom of the gravitational potential and that this potential does not evolve in time. Considering its position at the centre of mass of the Milky Way's nuclear cluster \citep[e.g.][]{feldmeier14}, it is highly unlikely that Sgr~A$^*$ by itself (i.e.~without the nuclear cluster) is moving with respect to the bottom of the global gravitational potential. Some motion of the central black hole is seen in large-scale numerical simulations due to the time-evolution of the gravitational potential, but this rapidly slows down once a nuclear cluster of only a few $10^5~\msun$ forms \citep{emsellem14}. The nuclear cluster of the Milky Way is $\sim2$ orders of magnitude more massive, yielding a combined mass of Sgr~A$^*$ and the nuclear cluster of several $10^7~\msun$, which dominates the gravitational potential out to $30~\pc$ \citep{launhardt02}. The energy required to move this entire structure relative to the global gravitational potential is substantial and may only be supplied by external perturbations such as large-scale instabilities, head-on (dwarf) galaxy mergers or encounters with other massive black holes. We therefore conclude that it is reasonable to fix Sgr~A$^*$ at the bottom of a time-invariant gravitational potential.

In view of these considerations, our orbital solution should become inaccurate if it is integrated for more than a single (azimuthal) orbital revolution in both directions. The part of the orbit considered in this paper falls well within that range. When integrating the orbit over a much longer time-scale, small deviations from the adopted gravitational potential would cause the model and real-Universe orbits to steadily diverge.

\subsubsection{The adopted geometry} \label{sec:geo}
We have assumed that the four identified gas streams can be fitted with a single orbit, running through the streams in the order Stream~2-3-4-1, where Streams~3 and~4 reside on the far side of the structure. There is no {\it a priori} reason to assume these are valid assumptions, but we justify them with a number of key observations. (1) The discontinuity in position-velocity space that is highlighted with the red circle in Figure~\ref{fig:orbit_mol} indicates that Streams~1 and~2 cannot be physically connected. (2) Streams~3 and~4 are coherent in the 3D phase space under consideration here. (3) While there could be a gap between Streams~4 and~1, they are easily connected in most orbital solutions without affecting the rest of the fit. (4) Likewise, we have omitted the widespread gas emission at higher longitudes from Sgr~B2 due to its complex kinematic structure and the resulting line-of-sight confusion. Its presence does suggest Streams~2 and~3 are connected, although it is unclear how far the orbit would extend. In order to maintain the same vertical oscillation period $P_z$, a larger apocentre radius would require a more strongly flattened potential (i.e.~a lower $q_{\Phi}$), which we show in Appendix~\ref{sec:potential} would yield unphysical mass distributions with negative densities. Finally, the emission at latitudes slightly higher than Sgr~C was assumed to belong to Stream~4. However, in our orbital model, this part of the gas stream is also indistinguishable from the beginning of Stream~2. This degeneracy cannot be lifted and hence the nature of this particular part of the gas stream remains ambiguous. This ambiguity is easily alleviated with future proper motion measurements of the Stream~4 clouds, because our model predicts that the proper motion vectors of Streams~2 and~4 have opposite directions. Until such measurements are available, we note that the quality of the fit is unaffected by the choice of geometry, because the gas fits both streams in our orbital model.

In summary, the best-fitting orbital structure is the simplest, physically motivated model that matches the observational constraints. More complex models (e.g.~fitting a larger number of independent streams) may yield better agreement with the observations, but do not necessarily lead to more physical insight.

\subsubsection{Relation to previous geometry estimates} \label{sec:geolit}
Previous work on the geometry of gas clouds in the CMZ provides independent constraints on the configuration implied by our orbital model. Here, we compare our model to these constraints and discuss the resulting implications for the local environment in which the CMZ clouds evolve and form stars.

It has been suggested that Sgr~B2 is located $130\pm60~\pc$ in front of Sgr~A$^*$ \citep{reid09}. This estimate relies on the assumption of a circular orbit. As shown in Table~\ref{tab:fit} and Figure~\ref{fig:orbit_td}, this assumption does not hold. Sgr~B2 has a total orbital velocity lower than the local circular velocity of the potential, because it follows an eccentric orbit and resides closer to apocentre than to pericentre. This places it at a distance of $\Delta y\sim38~\pc$ in front of Sgr~A$^*$, which in combination with the $\{l,b\}$ offset of $\Delta l=0.71^\circ$ between Sgr~B2 and Sgr~A yields the total galactocentric radius of $R=112^{+32}_{-9}~\pc$ listed in Table~\ref{tab:t}. Our estimate does agree with X-ray absorption studies. Figure~7 of \citet{ryu09} shows a line-of-sight separation between Sgr~B2 and Sgr~A$^*$ that is similar to the value reported here.

There exist different lines of indirect evidence suggesting that the $50$~and $20~\kms$ clouds are close to ($R\ll60~\pc$) or interacting with the gas in the circumnuclear disc (CND) orbiting Sgr~A$^*$. This is seemingly at odds with our orbital model, which has $R_{\rm p}=59^{+22}_{-19}~\pc$. The main evidence for a smaller pericentre radius is as follows.
\begin{enumerate}
\item
Sgr~A$^*$ resides within the shell of the supernova remnant Sgr~A~East, based on $90~{\rm cm}$ and OH~absorption measurements \citep{pedlar89,karlsson03}. Sgr~A~East has a diameter of $\sim10~\pc$. At the same time, the overabundance of OH maser emission at the positions where the supernova shell connects in projection to the $50~\kms$ cloud and the CND suggests the shell physically interacts with both \citep{yusefzadeh99,sjouwerman08,lee08}. If this indeed applies to the main body of the $50~\kms$ cloud, it would be situated within $\sim10~\pc$ of Sgr~A$^*$.
\item
The $50$~and $20~\kms$ clouds seem to form a contiguous structure in $\{l,b,v_{\rm los}\}$ space \citep[e.g.][]{sandqvist89}. As a result, the above point could subsequently imply that the $20~\kms$ is also located at a small distance from Sgr~A$^*$.
\item
Again in projection, there exist contiguous gas structures bridging the $50$~and $20~\kms$ clouds to the CND surrounding Sgr~A$^*$ \citep{herrnstein05,liu12,ott14}.
\end{enumerate}

Unfortunately, the above constraints on the positions of the {\it main bodies} of the $50$~and $20~\kms$ relative to Sgr~A$^*$ are rather qualitative or indirect. Both clouds are very close in projection to Sgr~A$^*$, where most of the orbital motion occurs in the plane of the sky. As a result, the line-of-sight velocities are small and most structures will be connected in $\{l,b,v_{\rm los}\}$ space irrespective of their distance to Sgr~A$^*$. The close projected distance between the $50$~and $20~\kms$ clouds and Sgr~A$^*$ complicates matters further by obstructing direct observations of any material that may be located in between. This is particularly important in the context of our model, because it is unknown how far the $50$~and $20~\kms$ clouds extend along the line of sight.

In our orbital model, the $50$~and $20~\kms$ clouds are very close to pericentre, indicating that they are likely undergoing tidal stripping. The stripped material from these and previously-passing clouds should make its way to the nucleus on a short ($R_{\rm p}/v_{\rm p}\sim0.3~\myr$) time-scale, where it should interact with Sgr~A~East and form a bridge between Sgr~A$^*$ and pericentre. The tidal perturbation during the pericentre passages of clouds on our best-fitting orbit could therefore provide a natural mechanism for feeding the CND. While different from the previously-proposed geometry, this configuration seems consistent with most of the above observational constraints. The numerical simulations that will be presented in Paper~II will allow us to address this in more detail.

\begin{figure}
\center\resizebox{\hsize}{!}{\includegraphics{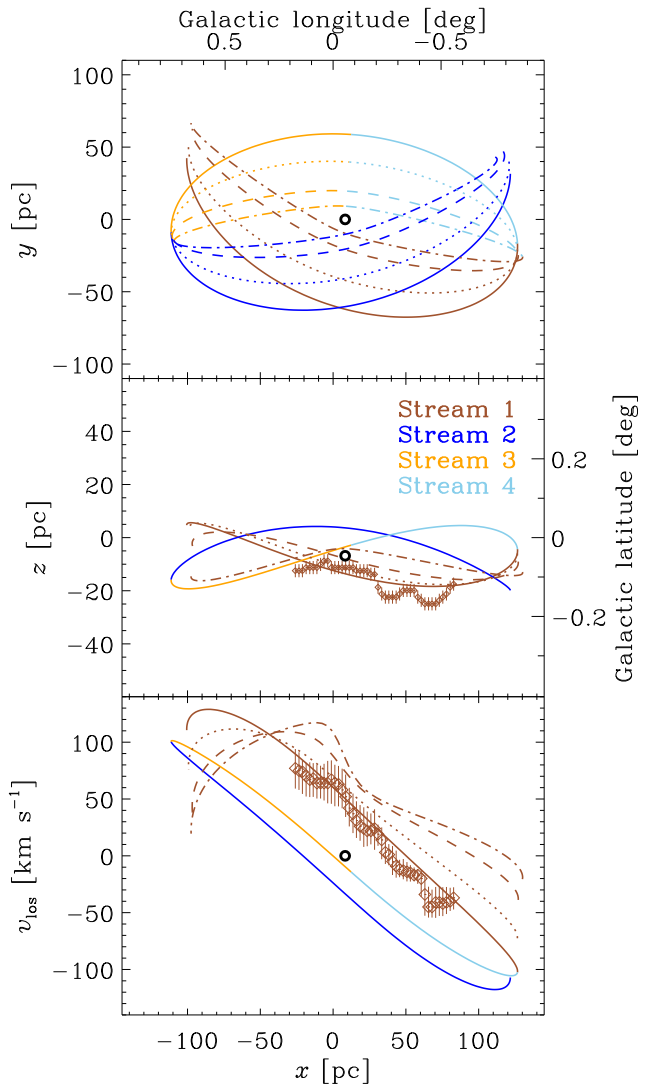}}\\
\caption[]{\label{fig:orbit_peri}
      {Comparison of the observed NH$_3(1,1)$ emission of Stream~1 with our orbital model (lines) for different pericentre radii $R_{\rm p}$. The \{dash-dotted, dashed, dotted, solid\} lines indicate pericentre radii of $R_{\rm p}=\{10,20,40,59\}~\pc$, where the latter value corresponds to our best-fitting model. As before, the colours refer to the four coherent streams in position-velocity space. The open black circle denotes the position of Sgr~A$^*$. {\it Top row}: Distribution in Galactic longitude and along the line of sight $\{x,y\}$. The different projected apocentre radii arise due to different heights above the plane. {\it Middle row}: Distribution in Galactic longitude and latitude $\{l,b\}$. Note the different scale on the $y$-axis compared to earlier figures. {\it Bottom row}: Distribution in Galactic longitude and line-of-sight velocity $\{l,v_{\rm los}\}$. The middle and bottom panels show the effect of a changing pericentre radius only for Stream~1.}
                 }
\end{figure}
Another constraint may be provided by differences in the relative stellar densities due to infrared absorption across the CMZ. For instance, the infrared stellar density observed in the dust ridge of clouds between the Brick and Sgr~B2 is lower than the $50$~and $20~\kms$ clouds. If this difference is due to absorption, it could suggest that the dust ridge clouds are positioned closer to the observer than the $50$~and $20~\kms$ clouds. However, this only holds if (1) the $50$~and $20~\kms$ clouds have column densities identical to the dust ridge clouds, and (2) the stellar densities just in front of the clouds are the same. Both of these requirements are questionable -- based on the radial distance of Table~\ref{tab:t}, we expect the stellar density at the positions of the $50$~and $20~\kms$ clouds to be twice as high as that in front of the dust ridge. In addition, the former clouds are projected against the brighter background of the nucleus. We thus see that in our model, the most prominent infrared dark clouds should lie on the dust ridge as observed.

While the above constraints may be qualitative, our orbital model can be used to quantify what the orbital kinematics should look like if the $50$~and $20~\kms$ clouds would indeed have a pericentre radius $R_{\rm p}\la10~\pc$. In Figure~\ref{fig:orbit_peri}, we show the effects of varying the pericentre radius on the best-fitting orbit (also see Appendix~\ref{sec:appendix}). As the pericentre radius decreases, the gradient of the line-of-sight velocity with longitude changes fundamentally. For orbits with large pericentre radii, we see that this gradient is monotonic, with a roughly constant slope in the $\{l,v_{\rm los}\}$ plane. However, orbits with pericentre radii $R_{\rm p}\leq40~\pc$ exhibit a rapid change of the line-of-sight velocity in the vicinity of Sgr~A$^*$. As the pericentre radius is decreased, the curves in the $\{l,v_{\rm los}\}$ plane become increasingly S-shaped, with near-constant $v_{\rm los}$ at large $|l|$ and a sudden jump at small $|l|$.

The rapid change of the line-of-sight velocity for small pericentre radii is robust -- irrespective of details such as the particular gravitational potential, the line-of-sight velocity near pericentre of any orbit changes fundamentally (i.e.~the sign changes or at least $|\Delta\ln{v_{\rm los}}/\Delta x|\sim1$) over a projected length-scale of one or two pericentre radii (i.e.~$\Delta x=1$--$2R_{\rm p}$). For $R_{\rm p}\sim10~\pc$ and an asymptotic line-of-sight velocity of $v_{\rm los}\sim100~\kms$, this implies a jump of $\Delta v_{\rm los}\sim60~\kms$ over a $\Delta x\sim20~\pc$ range in longitude, as is illustrated by the corresponding orbital model in Figure~\ref{fig:orbit_peri}.\footnote{{Also note that the model with $R_{\rm p}=10~\pc$ exhibits a strong change of direction in the $\{l,b\}$ plane near pericentre, continuing in the direction of negative latitudes.}} Such behaviour is inconsistent with the observed, near-constant slope of Stream~1 in the $\{l,v_{\rm los}\}$ plane across the full range of longitudes.

To verify if it is possible to obtain a smaller pericentre radius when allowing the other orbital parameters to vary, we have repeated the entire fitting procedure while fitting an orbital model to the data of Stream~1 {\it only}. In that case, we obtain best-fitting parameters typically within $1\sigma$ of those listed in Table~\ref{tab:fit}, with a pericentre radius of $R_{\rm p}=75^{+26}_{-35}~\pc$, even larger than the radius obtained when fitting all Streams ($R_{\rm p}=59^{+22}_{-19}~\pc$). Note that in both cases, the $1\sigma$ lower limit is $R_{\rm p,min}=40~\pc$. We therefore conclude that the only way the $50$~and $20~\kms$ clouds could be closer to Sgr~A$^*$ than $R_{\rm p}\sim40~\pc$, is if these clouds are unrelated to Stream~1. Such a disconnection seems unlikely given the strong coherence of Stream~1 across a large range in $\{l,b,v_{\rm los}\}$.

In summary, the geometry of the CMZ implied by our orbital model agrees with some of the geometries proposed in previous work, while disagreeing with others. However, the {\it observational constraints} on which these other geometries are based also seem to be consistent with our results, underlining the need for more quantitative and unambiguous constraints. We provide an example of such a quantitative constraint, showing that the line-of-sight velocities across Stream~1 are inconsistent with pericentre radii $R_{\rm p}\leq40~\pc$. This suggests that the $50$~and $20~\kms$ clouds may not be as close to Sgr~A$^*$ as previously thought.

\subsubsection{What is the origin of the asymmetry and gaps?} \label{sec:gaps}
We have fitted the orbital model to a gas distribution that contains several gaps where no NH$_3(1,1)$ emission is present. The distribution hosts two types of gaps. Firstly, there is a large-scale asymmetry in the CMZ where the vast majority of the gas emission comes from positive longitudes.\footnote{It is currently debated whether this asymmetry is mirrored by star formation. Most of the $24\mu{\rm m}$ emission (from young and evolved stars) is seen at negative longitudes where very little gas is present \citep[e.g.][]{yusefzadeh09}. This anti-correlation with the gas could be caused by the ambiguous origin of the $24\mu{\rm m}$ emission (i.e.~that many of the 24$\mu{\rm m}$ sources are evolved stars and therefore do not trace star formation, e.g.~\citealt{koepferl14}). Indeed, studies of young stellar objects, isolated massive stars, and young stellar clusters using Paschen-$\alpha$ or CO$_2$ ice absorption (which are not sensitive to evolved stars) find no such anti-correlation, but instead obtain a distribution that is similar to the lopsided gas distribution \citep[e.g.][]{mauerhan10,an11}.} A similar, associated asymmetry may be that the far side (Streams~3 and~4) of the structure appears to contain more tenuous gas than the front (Streams~1 and~2). Secondly, the gas streams themselves are interrupted by gaps of various sizes ($10$--$30~\pc$). Why does the gas emission not trace the orbit at all phase angles?

\citet{rodriguezfernandez08} show that a possible lopsidedness of the stellar potential is not responsible for the asymmetric gas distribution, as it would result in kinematics inconsistent with the observed line-of-sight velocities. Instead, these authors propose that accretion on to the CMZ may originate from only one side of the bar, which then enters the inner CMZ through the cloud complex at $l=1.3^\circ$. These asymmetries are commonly seen in external galaxies \citep[e.g.~NGC~5236, see][]{harris01b}.

While the above scenario could explain the large-scale, longitudinal asymmetry in the inner CMZ, it does not explain the asymmetry between the gas-rich front and gas-poor back side of the gas stream considered in this paper, because the front side must consist of two independent segments that in our orbital model represent the head and tail of the gas stream. It seems unlikely that enhanced gas densities at the two extreme ends of the same stream can be caused by asymmetric accretion. Even if Stream~1 is unrelated to Streams~2--4, this would require episodic accretion from {\it both} sides of the bar, contrary to the scenario of \citet{rodriguezfernandez08}. 

The density structure may be affected by the stream's orbital dynamics. The proximity to pericentre of the gas on Streams~1 and~2 should indeed lead to enhanced densities, but this explanation is incomplete -- Figure~\ref{fig:orbit_td} shows that the density should also peak near the third pericentre between Streams~3 and~4. In the context of our model, the only explanation is that the large-scale ($\sim100~\pc$) gas distribution along the gas stream is not contiguous, but clumpy. The triggered collapse of cloud complexes during the pericentre passages would then naturally lead to bursty star formation, consistent with the observed separation of gas overdensities and young stars (cf.~Figure~\ref{fig:orbit_td}). This would explain the absence of overdense gas near the pericentre between Streams~3 and~4, which in this picture coincides with a gap in the gas distribution.

There are two ways in which a clumpy large-scale gas distribution can be attained.
\begin{enumerate}
\item
The accretion flow on to the inner CMZ may be discontinuous. The associated length scale should be similar to the size-scale of the accretion shock. If the $l=1.3^\circ$ complex is the main accretion site of material on to the CMZ as suggested by \citet{rodriguezfernandez08}, then the corresponding size-scale is $\lambda\sim100~\pc$ \citep{kruijssen14b}.
\item
The gas in nuclear rings or streams within the inner Lindblad resonance (ILR; see \S\ref{sec:nature} below) develops gravitational instabilities of which the fastest growing mode has a wavelength of $\lambda\sim8.5\Delta R$, where $\Delta R$ is the stream thickness \citep{elmegreen94}. Substituting the observed $\Delta R\sim10~\pc$, we obtain $\lambda\sim85~\pc$.
\end{enumerate}
Both scenarios yield length-scales that are consistent with the implied separation of density enhancements ($\sim100~\pc$) in the large-scale asymmetry of the CMZ. More quantitative predictions require galaxy-scale simulations of the accretion process and the subsequent gravitational instabilities \citep[e.g.][]{emsellem14}.

Finally, the sizes of the small-scale ($10$--$30~\pc$) gaps in the gas stream are more easily understood. At the observed surface density and velocity dispersion of the gas stream \citep[mean values are $\Sigma\sim3\times10^3~\msun~\pc^2$ and $\sigma\sim15~\kms$, see][]{kruijssen14b}, the mean  turbulent Jeans length is $\lambda_{\rm J}=2\sigma^2/G\Sigma\sim35~\pc$, reaching $\lambda_{\rm J}\la20~\pc$ in overdensities like the Brick. This shows that the small-scale fragmentation of the gas stream naturally occurs (and leads to gaps) on size-scales consistent with the observed interruptions.

The small-scale gaps could be maintained under the influence of star formation and feedback. Star formation events in the gas stream are able to expel the gas locally, but their reach is limited. It therefore depends on the ratio between the separation length of star formation events ($\sim30~\pc$) and the feedback length scale (likely similar to the stream thickness of $\sim10~\pc$). For these numbers, we expect feedback to clear $\sim 30\%$ of the gas per star formation event, implying that the (interrupted) gas stream may survive for several pericentre passages, especially if material is reaccreted. This could explain why Stream~1 contains a substantial gas reservoir even though it represents the leading end of our model and may have experienced more than one pericentre passage in the past.\footnote{{In this context, it is important to reiterate the point made in \S\ref{sec:pred} that the gas properties of different parts of the stream differ. While a pericentre passage may induce collapse in one case, it could take several passages in another, depending on the density and velocity dispersion. This would also increase the longevity of the gas stream(s) in the CMZ.}} 

There is some tentative evidence that the proposed, {\it local} clearing of gas by feedback is presently ongoing. The molecular gas above (i.e.~the $80~\kms$ cloud at $\{l,b\}=\{0.1,0.2\}^\circ$) and below (i.e.~the tip of Stream~1) the Arches and Quintuplet clusters is connected to the bifurcation at the leading end of Stream~1 identified in \S\ref{sec:survey}. Combining the ages of the clusters ($\tau=3$--$5~\myr$, see \ref{sec:ymc}) and the half-separation length of these gas components above and below the gas stream ($R=20$--$30~\pc$), we obtain an ejection velocity of $\sim10~\kms$ for the dense gas shell(s), which is consistent with theoretical expectations \citep[cf.~Figures~1--3 of][]{murray10}.

Even where the gas stream does appear contiguous, imprints of the Jeans length should be present in the line-of-sight velocity profiles. This should manifest itself on a $\sim30~\pc$ length scale. We aim to address this in future work (Henshaw et al.,~in prep.).

\subsubsection{What is the nature of the stream(s)?} \label{sec:nature}
The three points discussed thus far in this section beg a more general question. What is the nature of the streams? Orbits in potentials generated by extended mass distributions are never closed in the inertial reference frame, but it is well known that barred potentials generate closed orbits in the rotating reference frame of the bar, which are often separated into a family of elongated `$x_1$' orbits along the bar and a family of perpendicular `$x_2$' orbits embedded within the $x_1$ orbits \citep[e.g.][]{contopoulos77,binney91,athanassoula92,sellwood93,englmaier99,bissantz03}. Because these orbits are closed in the rotating reference frame of the bar, they are open by definition in the Galactic reference frame, with orbital precession rates matching the bar's angular speed. Closed orbits are often required for the gas to avoid self-interaction and hence disruption, but we note that this is not required if there is a non-negligible vertical oscillation (like in our best-fitting orbit). In such a case, it takes several orbits before the gas streams cross and interact.

Could the gas stream and its best-fitting orbit be consistent with the $x_2$ orbits? This was first proposed by \citet{binney91}, who used low-density gas tracers to characterise the gas dynamics. We revisit the question here using our orbital fit to high-density gas tracers. A wide range of $x_2$ ring radii has been measured in external galaxies. While most of these extend beyond the size-scales of the gas stream considered here, several of them are similar in size \citep[e.g.~NGC~1068, see][]{peeples06}. Most $x_2$ orbits reside just interior to the ILR \citep[e.g.][]{regan03}, but the innermost orbits extend to smaller radii. For instance, the ILR in the Milky Way model of \citet[Figure~10]{bissantz03} resides at a radius of $R\sim200~\pc$, whereas the innermost $x_2$ orbits reach $R\sim20~\pc$.

There is circumstantial observational evidence that our best-fitting orbit may coincide with a resonance. For instance, the orientation of the line connecting both apocentres is closer to being perpendicular to the orientation of the bar than running in parallel to it. The ILR occurs at the galactocentric radius where $\Omega-\kappa/2=\Omega_{\rm p}$, with $\Omega$ the angular velocity, $\kappa$ the epicyclic frequency, and $\Omega_{\rm p}$ the pattern speed of the bar. In the Milky Way, the ILR is thought to reside at radii beyond the fitted orbit \citep[e.g.][]{englmaier99}, but the above {\it range} of size-scales shows that it is possible that our orbit matches the Galactic $x_2$ orbits. We can test the hypothesis by comparing the orbital rate of precession $\Omega_{\rm prec}$ to the pattern speed of the bar $\Omega_{\rm p}\sim0.06~\myr^{-1}$ \citep[e.g.][]{debattista02,bissantz03,gardner10}. If these angular velocities match, then our fitted orbit is closed in the reference frame of the bar.

The precession rate of the best-fitting orbit is given by $\Omega_{\rm prec}=2\pi/P_\phi-\pi/P_R$, where the absence of the factor of 2 in the second term arises because gas on the $x_2$ orbits experiences two peri/apocentre passages per orbital revolution. In the reference frame of the bar, the precession rate becomes $\hat{\Omega}_{\rm prec}=\Omega_{\rm prec}-\Omega_{\rm p}$. Because $x_2$ orbits are closed, they must have $\hat{\Omega}_{\rm prec}=0$ by definition. Using the orbital periods from Table~\ref{tab:fit} and accounting for their covariance, we obtain $\Omega_{\rm prec}=0.16^{+0.13}_{-0.01}~\myr^{-1}$ and hence $\hat{\Omega}_{\rm prec}=0.10^{+0.13}_{-0.01}~\myr^{-1}$. This indicates that our best-fitting orbit is inconsistent with the $x_2$ orbits. The reason for this inconsistency can be inferred directly from the potential implied by the mass profile of \citet{launhardt02}. In the radial range of $R=0$--$300~\pc$ under consideration here, the precession rate $\Omega-\kappa/2$ has a minimum of $0.13~\myr^{-1}$ at $R=110~\pc$, twice as high as the pattern speed of the bar. The condition for closed orbits is thus satisfied nowhere in this radial range. Extrapolating the mass distribution of \citet{launhardt02} to larger radii suggests that closed orbits exist in the radial range $R=300$--$700~\pc$, well outside the range of orbital solutions considered here.

Despite this clear inconsistency between our best-fitting solution and the $x_2$ orbits, we caution against drawing firm conclusions from this comparison. The difference between $\Omega_{\rm prec}$ and $\Omega_{\rm p}$ is hardly significant given the systematic uncertainties involved, such as the possible deviations from axisymmetry discussed in \S\ref{sec:pot}, which could provide the torque necessary to decrease the precession rate and close the orbit in the rotating reference frame of the bar. The gas stream's kinematics may also be affected by viscous forces, which were neglected in the dynamical model presented here.\footnote{While we acknowledge the possibility, we note that deviations from our ballistic orbital model due to hydrodynamics require the stream to consistently encounter gas of similar (or higher) density. Because the best-fitting orbit rarely intersects with itself (see \S\ref{sec:clouds}) and the dense gas in the CMZ has a low volume filling factor \citep{longmore13}, hydrodynamical perturbations are likely rare too. Our assumption of ballistic dynamics is therefore reasonable.} We therefore cannot rule out that the gas stream formed due to the $x_2$ resonance.

Alternatively, observations of external galaxies often reveal several (sometimes point-symmetric) elongated `feathers' that emerge from the inside of the $x_2$ orbits, reaching in to the small radii where the nuclear clusters and the central black hole dominate the gravitational potential \citep{peeples06}. Like a `closed' $x_2$ orbit, these may also have precession rates similar to the bar, but even then their kinematics should be fundamentally different. The position-velocity distribution of the streams identified in the CMZ is not point-symmetric, but may none the less be consistent with the feather hypothesis. Examples of deviations from point-symmetry are not uncommon in extragalactic systems (see e.g.~NGC~1097 and NGC~6951 in the sample of \citealt{peeples06}), where the feathers continue to orbit the galaxy centre on eccentric orbits similar to what we see in the CMZ. Such kinematics have also been found in the recent disc galaxy simulation by \citet{emsellem14}. We therefore emphasise the possibility that the identified streams may represent the Galactic analogue of the feathers seen in extragalactic observations. Considering that orbits similar to our model should exist in any vertically-compressed, extended mass distribution, this is an interesting avenue for future high-resolution observations of gas streams in external galaxy centres (e.g.~using ALMA).

\subsubsection{The relation to the Arches and Quintuplet clusters} \label{sec:ymc}
The CMZ hosts the Arches and Quintuplet clusters, which are the only two known young ($\tau<10~\myr$) massive ($M\ga10^4~\msun$) clusters in the region. Did these clusters form from gas following our orbital model? About $50\%$ of the star formation in the CMZ is thought to occur in bound clusters as shown by observations \citep{mauerhan10} and theory \citep{kruijssen12d}, of which the majority is destroyed on time-scales of $\sim10~\myr$ \citep[see e.g.][]{portegieszwart01,kruijssen11,kruijssen14b}. It is plausible that the Arches and Quintuplet clusters represent the high-mass end of this cluster population.

The Arches has a line-of-sight velocity of $v_{\rm los}=109\pm8~\kms$ in the Galactic reference frame \citep{figer02} and a proper motion of $v_{\rm pm}=172\pm15~\kms$ with respect to the background field stellar population, almost entirely in the Galactic plane towards increasing longitudes \citep{clarkson12}.\footnote{This number is a downward revision from \citet{stolte08}.} Together, this implies a 3D space velocity of $v_{\rm orb}=204\pm13~\kms$. For the Quintuplet cluster, we obtain a similar result -- its line-of-sight velocity is $v_{\rm los}=116\pm2~\kms$ in the Galactic reference frame, whereas its proper motion with respect to the field stellar population is $v_{\rm pm}=132\pm15~\kms$, again almost entirely in the positive longitude direction \citep{liermann09,stolte14}, implying a 3D space velocity of $v_{\rm orb}=176\pm15~\kms$.

\citet{stolte08} compared the line-of-sight velocities of the Arches and Quintuplet clusters to those of the gas stream and argued that the clusters must follow different orbits than the gas, suggesting they were formed by cloud-cloud collisions. However, our model shows that the high eccentricity of the orbit allows a wide range of line-of-sight velocities for different projection angles of the velocity vectors, depending on where along the orbit the object in question is located. The mean 3D velocity vectors of both clusters are constituted by line-of-sight components $\langle v_{\rm los}\rangle=113\pm4~\kms$ in the Galactic reference frame and $\langle v_{\rm los}'\rangle=99\pm4~\kms$ in the local standard of rest, as well as a mean 2D proper motion towards positive longitudes of $\langle v_{\rm pm}\rangle=152\pm21~\kms$. Comparing to our complete orbital solution in Appendix~\ref{sec:fullorbit} at the time when Stream~1 best matches the observed $\{l,b\}$ coordinates of both clusters ($t=2.1~\myr$), we see that the predicted line-of-sight and 2D proper motion velocities are $v_{\rm los}=91~\kms$ and $v_{\rm pm}=185~\kms$, respectively, both in reasonable agreement with the observed values. As a result, the orbital velocities of both clusters (with mean $\langle v_{\rm orb}\rangle=190\pm20~\kms$) are also fully consistent with our modelled orbital velocity at that position ($v_{\rm orb}=206~\kms$). These velocities show that the clusters are consistent with being part of Stream~1 in our orbital model, very close to pericentre. If both clusters indeed follow our best-fitting orbit, they should presently reside at a galactocentric radius of $R\sim60~\pc$.

\begin{figure}
\center\resizebox{\hsize}{!}{\includegraphics{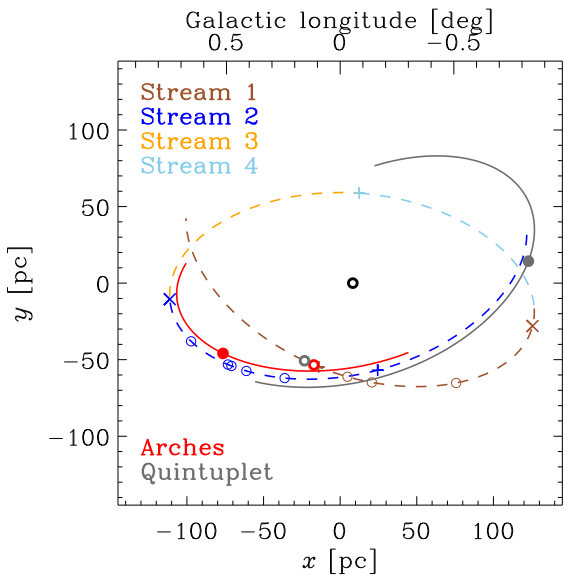}}\\
\caption[]{\label{fig:orbit_cl}
      {Top-down view of the present and past of the Arches and Quintuplet clusters in the context of our orbital model (dashed line). As in Figure~\ref{fig:orbit_td}, the observer is located in the negative-$y$ direction. The colours again refer to the four coherent streams in position-velocity space, the thin circles indicate the implied positions in the Galactic plane of several GMCs and cloud complexes in the CMZ, the plus symbols indicate pericentres, the crosses mark apocentres, and the open black circle denotes the position of Sgr~A$^*$. The present-day positions of the Arches and Quintuplet clusters are represented by the thick circles on Stream~1, close to pericentre. The formation sites of these clusters as implied by their ages are indicated by the solid dots on lines running in parallel to the orbital model. The lengths of the lines indicate the uncertainty ranges implied by the age measurements.}
                 }
\end{figure}
The {\it present} configuration of the gas does not provide much insight into the formation sites of the Arches and Quintuplet, but its dynamical history may. Figure~\ref{fig:orbit_cl} shows the implied present-day positions of the Arches and Quintuplet clusters in our orbital model, as well as the possible range of their formation sites implied by our model. The clusters have ages of $\tau=3.5\pm0.7~\myr$ and $\tau=4.8\pm1.1~\myr$, respectively \citep{schneider14}. In our orbital model, these ages indicate that the clusters are ahead of their formation sites by $0.9\pm0.2$ and $1.3\pm0.3$ full orbits (i.e.~azimuthal periods), or $1.7\pm0.5$ and $2.4\pm0.7$ radial oscillations. We see that the ages of both clusters are consistent with an integer number of radial oscillations. If their present positions are indeed near the pericentre passage of Stream~1 as our model suggests, then the clusters likely formed near the pericentre passage of Stream~2, after which they completed approximately one orbital revolution to end up at their present-day positions. The range of their possible formation sites is indicated in Figure~\ref{fig:orbit_cl} by the solid lines, which show that the uncertainties are substantial due to the large error bars on the age measurements. While the Quintuplet could have formed at any point of a complete radial oscillation, the Arches is very much consistent with having formed in the dust ridge between the Brick and Sgr~B2. In our model, its formation was triggered by the tidal compression of clouds during the preceding pericentre passage.

The currently available evidence supports the scenario that the Arches and Quintuplet clusters formed in the gas stream, but uncertainties remain. The age estimates represent the main source of uncertainty \citep[cf.][]{figer99,najarro04,schneider14}, but decreasing the error margins on our orbital fit in future work (e.g.~using a better-constrained gravitational potential) could also improve the above analysis.

\subsection{Implications and outlook}
The presented orbital model provides a robust starting point for observational, theoretical and numerical follow-up studies of GMC evolution in the inner CMZ. Observationally, the predictions of \S\ref{sec:pred} and the open questions of \S\ref{sec:questions} can be addressed using the plethora of radio, sub-mm and infrared survey data that is already at hand. In addition, our assumptions can be improved upon by refining our current understanding of the gravitational potential in the inner $200~\pc$ of the Milky Way. 

Perhaps most importantly, our model of an absolute time-sequence of GMC evolution provides quantitative constraints that will aid the interpretation of upcoming, high-resolution observations of these clouds (e.g.~using ALMA). The obvious next step is to follow the time-evolution along the orbit of several processes that govern cloud evolution and star formation, such as the turbulent energy dissipation through shocks, fragmentation into cores, star formation activity, and the distribution of gas temperatures, volume densities, chemistry, and magnetic field strengths.

In Paper~II, we will present hydrodynamical simulations of gas clouds that are orbiting the Galactic Centre on the best-fitting orbit presented in this work. With these simulations, we aim to investigate the structure and dynamics of the observed clouds, paying particular attention to the influence of the pericentre passage on the cloud properties. This will provide a wide range of quantitative predictions that can be tested with the observations outlined above.

\section*{Acknowledgements}
This work has benefited from insightful discussions with John Bally, Michael Burton, Eric Emsellem, Chervin Laporte, Adriane Liermann, Sergio Molinari, Jill Rathborne, Andrea Stolte, Leonardo Testi, and Simon White. We thank Michael Burton, Adam Ginsburg, Katharina Immer, Katharine Johnston, Witold Maciejewski, Betsy Mills, and J\"{u}rgen Ott for helpful comments that improved the paper, as well as Ralf Launhardt for providing the scripts used in generating the figures of \citet{launhardt02}. JED acknowledges support from the DFG cluster of excellence `Origin and Structure of the Universe'. The image of the dragon in Figure~\ref{fig:betaq} was adapted from a facsimile of Pierre Descelier's 1546 world map by E.~Rembielinski (published 1862), National Library of Australia, MAP~RM~567. This work made use of data from the Midcourse Space Experiment. Processing of the data was funded by the Ballistic Missile Defense Organization with additional support from NASA Office of Space Science. This research has also made use of the NASA/IPAC Infrared Science Archive, which is operated by the Jet Propulsion Laboratory, California Institute of Technology, under contract with the National Aeronautics and Space Administration.

\appendix

\bibliographystyle{mn2e}
\bibliography{mybib}

\section{Adopted gravitational potential} \label{sec:potential}
The gravitational potential in the central few hundred~pc of the CMZ is dominated by stellar mass. The spherically-symmetric, enclosed mass distribution was derived by \citet{launhardt02},\footnote{\citet{launhardt02} assumed a distance to the Galactic Centre of $R_{\rm L02}=8.5~\kpc$. Since we adopt a distance of $R=8.3~\kpc$, we rescale the radii by a factor of $R/R_{\rm L02}$ and the enclosed masses by a factor of $(R/R_{\rm L02})^2$.} which averaged over the range $R=1$--$300~\pc$ results in a density profile $\rho(R)\propto R^{-\gamma}$ with $\gamma=1.7$--$1.9$.

\begin{figure}
\center\resizebox{\hsize}{!}{\includegraphics{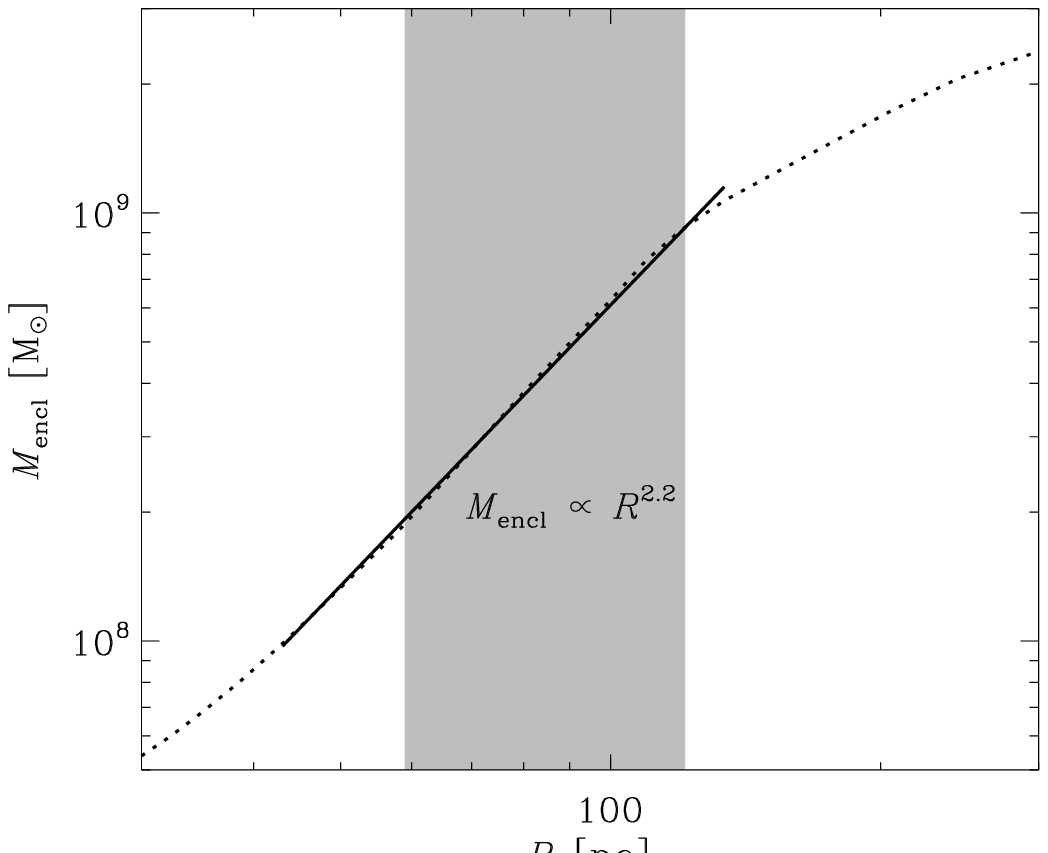}}\\
\caption[]{\label{fig:mencl}
      {Enclosed mass as a function of galactocentric radius in the range $R=30$--$300~\pc$ from \citet[dotted line]{launhardt02}. The grey region shows the radial range spanned by the best-fitting pericentre and apocentre distances, whereas the solid line shows a power-law fit to the enclosed mass profile in the radial range spanned by our best-fitting pericentre (minus one standard deviation) and apocentre (plus one standard deviation).}
                 }
\end{figure}
In order to reproduce the vertical oscillations of the stream's orbit described in \S\ref{sec:improv}, we assume that the gravitational potential in the Galactic Centre is axisymmetric and flattened. Ideally, the vertical compression should be performed on the underlying mass density distribution, but due to the observational resolution, there is insufficient information on the true vertical density profile at the small latitudes ($|z|<15~\pc$) considered here \citep[cf.][]{launhardt02}. We therefore flatten the gravitational potential itself and show below that doing so yields physically-allowed potential-density pairs. A second, possible concern is that the potential is the sum of several components, including those from Sgr~A$^*$ and the nuclear stellar cluster. These latter two components are not flattened in reality, but they represent only a small fraction ($\sim10\%$) of the total potential in the radial range occupied by our best-fitting orbit. For simplicity, we therefore construct a single, flattened potential.

The potential is flattened by the coordinate transformation
\be
\Phi(R)\rightarrow\Phi(r,z),
\ee
where
\be
R^{2}\equiv r^{2}+\frac{z^{2}}{q_{\Phi}^{2}},
\ee
with $R$ the 3D radius, $r\equiv(x^2+y^2)^{1/2}$ the 2D radius in the Galactic plane, $z$ the height above the plane, and $q_{\Phi}\leq1$ a free parameter describing the degree of flattening. A spherically symmetric potential is described by $q_\Phi=1$.

Vertically compressing the potential through the above coordinate transformation can yield (locally) negative densities if either the flattening is too strong or the density profile too steep \citep{binney87}. Given an initially spherically-symmetric, power-law density profile $\rho\propto R^{-\gamma}$ (and hence $M_{\rm encl}\propto R^{\beta}$ with $\beta=3-\gamma$), it can be shown that negative densities do not occur if the flattening parameter $q_\Phi$ obeys
\be
q_{\Phi}<(1-\beta)^{-\frac{1}{2}},
\ee
which is always satisfied if $\beta>1$, and
\be
q_{\Phi}>(1-\beta/2)^{\frac{1}{2}},
\ee
which is always satisfied if $\beta>2$. The slope of the enclosed mass profile is thus critical in determining whether the flattening of the gravitational potential yields a physically-allowed density distribution. In Figure \ref{fig:mencl}, we show the enclosed mass as a function of the galactocentric radius in the range $R=30$--$300~\pc$ from \citet{launhardt02}, as well as a power-law fit in the radial range spanned by our best-fitting orbital solution(s). This gives a best-fitting slope of $\beta=2.2$ over the full radial range. Locally, the value of $\beta$ across the same radial interval ranges from $\beta=1.4$ to $\beta=2.4$, implying that flattening parameters of $0.5\la q_\Phi\leq1$ result in physically-allowed density distributions at all radii.

\begin{figure}
\center\resizebox{\hsize}{!}{\includegraphics{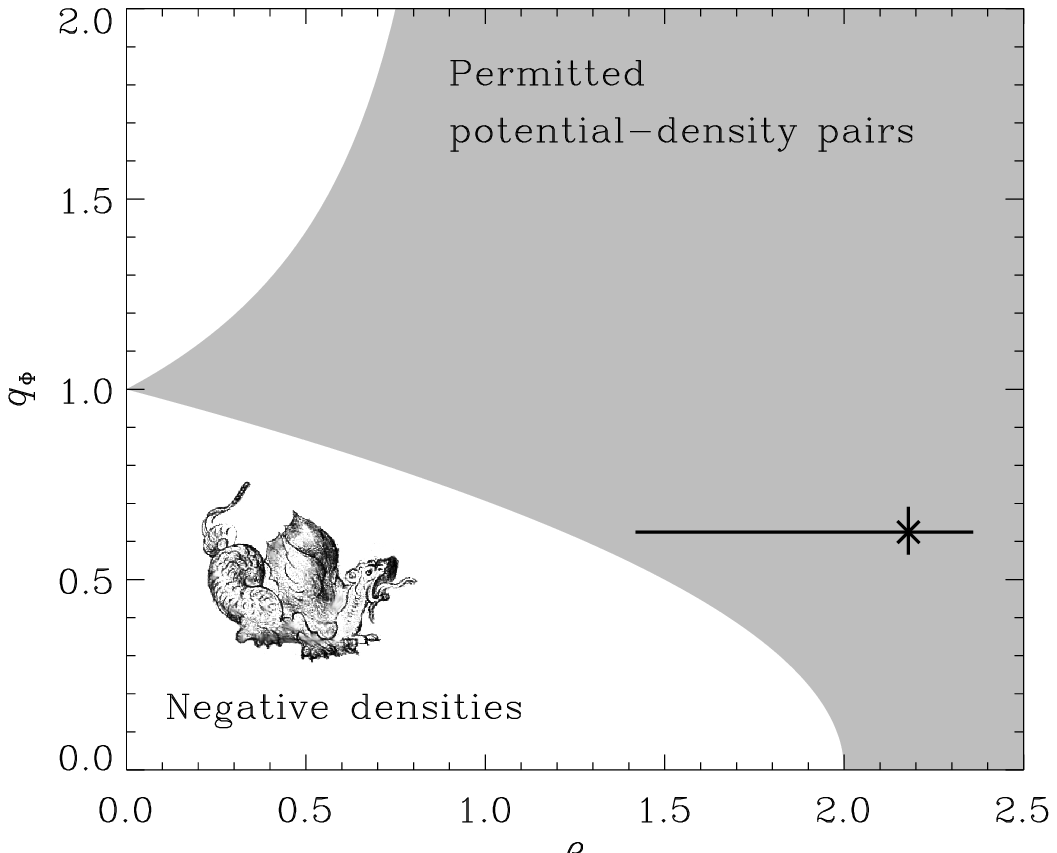}}\\
\caption[]{\label{fig:betaq}
      {Parameter space spanned by the enclosed mass profile slope $\beta$ and the potential flattening parameter $q_\Phi$. The region in which the flattened potential corresponds to physically-allowed density distributions is shaded in grey. The cross represents the best-fitting enclosed mass slope $\beta=2.2$ and the best-fitting potential flattening parameter $q_\Phi=0.63$. The vertical error bar denotes the formal uncertainty on $q_{\Phi}$, whereas the horizontal error bar represents the range of $\beta$ across the fitted radial interval in Figure~\ref{fig:mencl}.}
                 }
\end{figure}

Having determined the slope of the enclosed mass profile in the region of interest, we show the $\{\beta, q_{\Phi}\}$ parameter space in Figure \ref{fig:betaq} with the permitted region shaded in grey. The cross symbol marks the position defined by our best-fitting slope $\beta$ and the best-fitting potential flattening parameter $q_\Phi$ (see Table~\ref{tab:fit}). Note that the horizontal error bar represents the range of $\beta$ across the fitted radial interval in Figure~\ref{fig:mencl} (i.e.~{\it not} the standard deviation). We see that flattening the gravitational potential implied by the enclosed mass profile from \citet{launhardt02} yields physically-allowed density distributions for any combination of the parameters $\beta$ and $q_{\Phi}$ considered in this work.

\section{The dependence of the best-fitting orbit on the orbital parameters} \label{sec:appendix}
\begin{figure*}
\center\resizebox{165mm}{!}{\includegraphics{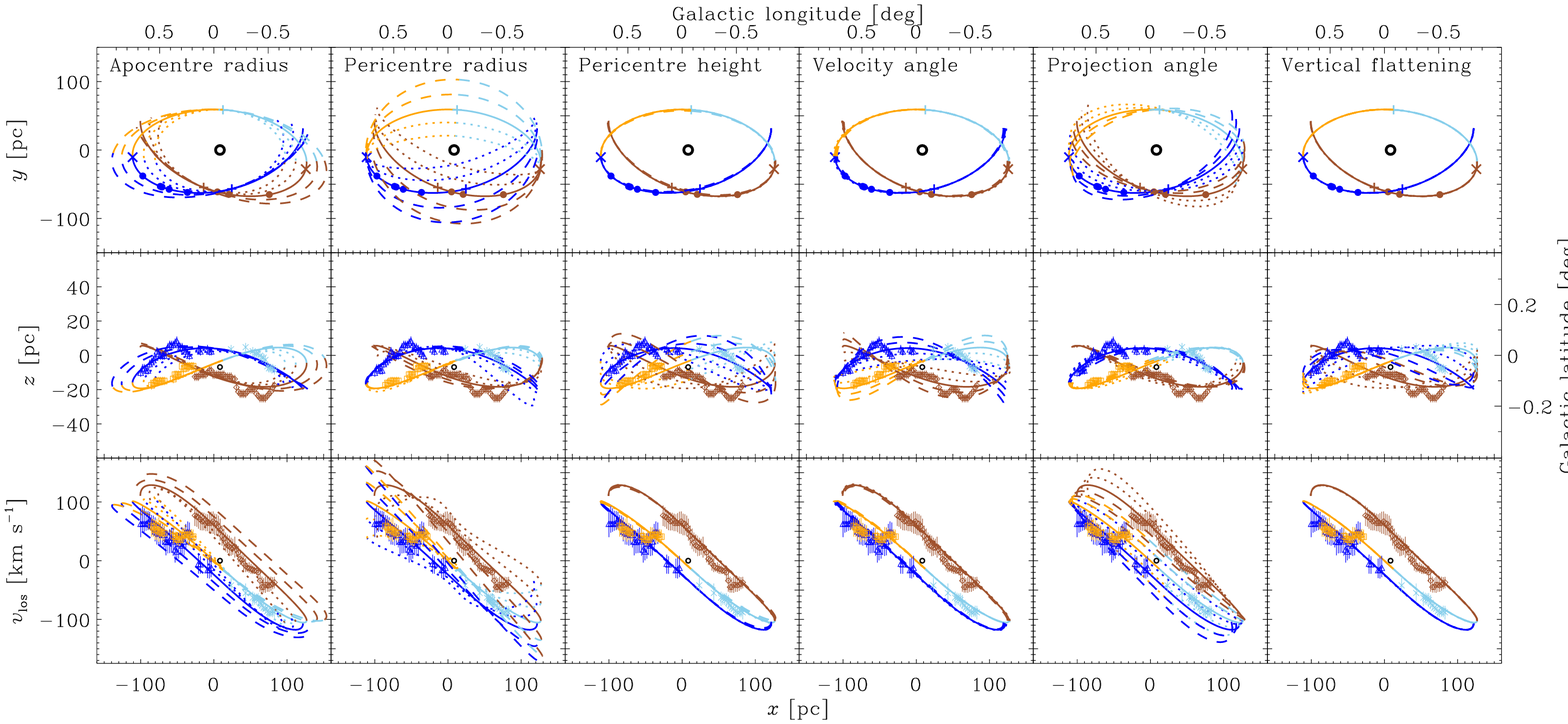}}\\
\caption[]{\label{fig:fit}
      {Comparison of the observed NH$_3(1,1)$ emission (symbols with error bars, tracing gas with volume densities $n>{\rm several}~10^3~\cmc$) near the Galactic Centre with our orbital model (solid line). The two dotted (dashed) lines indicate the effect of decreasing (increasing) the parameter indicated at the top left of each column by $1\sigma$ and $2\sigma$ (see Table~\ref{tab:fit}). As in Figures~\ref{fig:orbit_mol},~\ref{fig:orbit_fit} and~\ref{fig:orbit_td}, the colours refer to the four coherent streams in position-velocity space. The open black circle denotes the position of Sgr~A$^*$. {\it Top row}: Distribution in Galactic longitude and along the line of sight $\{x,y\}$. Dots are positioned along the best-fitting orbital model to indicate the longitudes of the clouds discussed in \S\ref{sec:clouds} and Figure~\ref{fig:orbit_td}, whereas crosses (plus symbols) mark the apocentres (pericentres). {\it Middle row}: Distribution in Galactic longitude and latitude $\{l,b\}$. {\it Bottom row}: Distribution in Galactic longitude and line-of-sight velocity $\{l,v_{\rm los}\}$.}
                 }
\end{figure*}
Figure~\ref{fig:fit} shows the dependence of the best-fitting orbit on the six free parameters that we used. It shows that the $\{l,b\}$ distribution is affected by all parameters, but the $\{l,v_{\rm los}\}$ distribution is only affected by varying $R_{\rm a}$, $R_{\rm p}$ and $\phi$. This is easily understood -- as explained in \S\ref{sec:orbit}, the best-fitting orbit is integrated around the far-side pericentre between Streams~3 and~4 (cf.~Figure~\ref{fig:orbit_td}). Because the far-side pericentre lies almost exactly along the line observer--Sgr~A$^*$ (i.e.~the projection angle is $\phi\sim180^\circ$), the parameters $z_{\rm p}$, $\theta$ and $q_{\Phi}$ only affect the orbit in directions perpendicular to the line of sight, leaving $v_{\rm los}$ unaffected.

The figure clearly demonstrates that the parameters are non-degenerate. The three variables that set the $\{l,v_{\rm los}\}$ distribution do so in distinctive ways, which is illustrated most clearly by observing the variation of the extrema of the line-of-sight velocity in the bottom panels of Figure~\ref{fig:fit}. The apocentre radius $R_{\rm a}$ mainly affects (1) the velocity range spanned by the stream segments at a given Galactic longitude and (2) the longitude at which the peak $v_{\rm los}$ is reached, without strongly influencing its value. The opposite behaviour is seen when varying the pericentre radius $R_{\rm p}$ -- the peak velocity changes substantially while the extrema in Galactic longitude does not vary as much, thereby changing the slopes of the curves. Finally, changing the projection angle $\phi$ mainly changes the line-of-sight velocity normalisation of the system at a roughly constant slope.

Focussing on the configuration in $\{l,b\}$ space, the effects of the apocentre and pericentre radii are similar as in $\{l,v_{\rm los}\}$ space, i.e.~the apocentre radius affects the orbit's extent in Galactic longitude, whereas the changing the pericentre radius results in modest differences within a fixed longitude range. The pericentre height $z_{\rm p}$ affects the Galactic latitude of the orbit at all longitudes. Most notably, the latitude shifts at extreme longitudes are opposite to those near the projected centre of the orbit. While the velocity angle $\theta$ appears to have a similar effect on the latitude, a closer inspection shows its effect at extreme longitudes does not mirror that at the centre. Instead, it introduces a vertical stretch or compression at all longitudes, while having the opposite effect on the orbit's extent in Galactic latitude. This is not surprising, because the velocity angle determines which fraction of the kinetic energy is used for the vertical and radial oscillations. The projection angle $\phi$ only weakly affects the structure in $\{l,b\}$ space, in clear contrast with its effect on $\{l,v_{\rm los}\}$ space that was discussed above. It shifts the extrema in Galactic longitude, but does so asymmetrically -- an increase of the low-longitude extremum is accompanied by a decrease of the high-longitude extremum and vice versa. This mirrors the effect of changing the apocentre radius, which does symmetrically extend or compress the Galactic longitude range spanned by the orbit. Finally, the vertical flattening of the potential $q_{\Phi}$ changes the extreme-longitude ends of the orbit in a way similar to the pericentre height, but does not affect the far side of the orbit (Streams~3 and~4) like $z_{\rm p}$ does.

In summary, each of the six parameters has its own unique effect on the structure of the orbit in $\{l,b,v_{\rm los}\}$ space. For this reason it is possible to obtain a reliable, non-degenerate orbital fit.

\section{The complete orbital solution} \label{sec:fullorbit}
\begin{table*}
 \centering
  \begin{minipage}{180mm}
  \caption{Complete orbital solution}\label{tab:fullorbit}
  \begin{tabular}{@{}c c c c c c c c c c c c c c c c@{}}
  \hline 
  $t$ & $x$ & $y$ & $z$ & $R$ & $v_x$ & $v_y$ & $v_z$ & $v_{\rm orb}$ & $l$ & $b$ & $v_{\rm los}'$ & $\mu_l'$ & $\mu_b'$ & $\mu_x'$ & $\mu_y'$ \\
  (1) & (2) & (3) & (4) & (5) & (6) & (7) & (8) & (9) & (10) & (11) & (12) & (13) & (14) & (15) & (16) \\
  \hline
  -2.5 & 100.90 & -13.09 & -12.04 & 93.69 & -106.02 & -115.60 & 31.17 & 159.93 & -0.680 & -0.081 & -129.60 & -3.75 & 0.57 & -2.45 & -2.89 \\
  -2.4 & 88.71 & -24.65 & -8.69 & 84.15 & -132.19 & -109.83 & 33.88 & 175.17 & -0.598 & -0.059 & -123.83 & -3.10 & 0.64 & -2.17 & -2.31 \\
  -2.3 & 73.93 & -35.41 & -5.20 & 74.62 & -156.21 & -99.85 & 33.92 & 188.48 & -0.498 & -0.035 & -113.85 & -2.50 & 0.64 & -1.85 & -1.80 \\
  -2.2 & 56.88 & -44.92 & -1.86 & 66.38 & -176.55 & -85.41 & 30.82 & 198.54 & -0.383 & -0.013 & -99.41 & -2.00 & 0.56 & -1.52 & -1.41 \\
  -2.1 & 38.01 & -52.73 & 0.99 & 61.06 & -191.47 & -66.65 & 24.47 & 204.21 & -0.256 & 0.007 & -80.65 & -1.63 & 0.41 & -1.20 & -1.18 \\
  -2.0 & 17.97 & -58.44 & 3.05 & 60.06 & -199.24 & -44.51 & 15.38 & 204.73 & -0.121 & 0.021 & -58.51 & -1.43 & 0.18 & -0.90 & -1.13 \\
  -1.9 & -2.46 & -61.79 & 4.09 & 63.67 & -198.97 & -20.98 & 4.89 & 200.14 & 0.017 & 0.028 & -34.98 & -1.44 & -0.08 & -0.68 & -1.27 \\
  -1.8 & -22.47 & -62.74 & 4.05 & 70.73 & -191.28 & 2.00 & -5.46 & 191.38 & 0.151 & 0.027 & -12.00 & -1.63 & -0.34 & -0.56 & -1.57 \\
  -1.7 & -41.37 & -61.44 & 3.01 & 79.61 & -177.51 & 23.24 & -14.59 & 179.62 & 0.279 & 0.020 & 9.24 & -1.97 & -0.56 & -0.55 & -1.98 \\
  -1.6 & -58.62 & -58.08 & 1.14 & 88.95 & -159.12 & 42.09 & -21.81 & 166.04 & 0.395 & 0.008 & 28.09 & -2.43 & -0.74 & -0.63 & -2.46 \\
  -1.5 & -73.80 & -52.92 & -1.37 & 97.82 & -137.42 & 58.24 & -26.83 & 151.65 & 0.497 & -0.009 & 44.24 & -2.97 & -0.87 & -0.80 & -2.99 \\
  -1.4 & -86.64 & -46.25 & -4.28 & 105.62 & -113.29 & 71.73 & -29.62 & 137.34 & 0.584 & -0.029 & 57.73 & -3.57 & -0.94 & -1.05 & -3.54 \\
  -1.3 & -96.91 & -38.33 & -7.36 & 111.96 & -87.27 & 82.71 & -30.29 & 124.01 & 0.653 & -0.050 & 68.71 & -4.21 & -0.95 & -1.37 & -4.10 \\
  -1.2 & -104.47 & -29.43 & -10.40 & 116.58 & -60.40 & 91.06 & -28.99 & 113.05 & 0.704 & -0.070 & 77.06 & -4.88 & -0.92 & -1.74 & -4.65 \\
  -1.1 & -109.25 & -19.80 & -13.23 & 119.35 & -33.06 & 96.88 & -25.94 & 105.62 & 0.736 & -0.089 & 82.88 & -5.56 & -0.85 & -2.15 & -5.19 \\
  -1.0 & -111.21 & -9.70 & -15.66 & 120.21 & -5.43 & 100.29 & -21.37 & 102.72 & 0.750 & -0.106 & 86.29 & -6.24 & -0.73 & -2.60 & -5.72 \\
  -0.9 & -110.35 & 0.63 & -17.55 & 119.11 & 22.36 & 101.33 & -15.51 & 104.94 & 0.744 & -0.118 & 87.33 & -6.93 & -0.59 & -3.09 & -6.24 \\
  -0.8 & -106.65 & 10.95 & -18.79 & 116.06 & 50.19 & 99.92 & -8.56 & 112.15 & 0.719 & -0.127 & 85.92 & -7.62 & -0.41 & -3.59 & -6.74 \\
  -0.7 & -100.09 & 20.98 & -19.27 & 111.08 & 77.92 & 95.91 & -0.76 & 123.59 & 0.675 & -0.130 & 81.91 & -8.31 & -0.22 & -4.11 & -7.23 \\
  -0.6 & -90.72 & 30.46 & -18.93 & 104.28 & 105.08 & 89.13 & 7.50 & 138.01 & 0.612 & -0.128 & 75.13 & -8.99 & -0.02 & -4.64 & -7.70 \\
  -0.5 & -78.65 & 39.11 & -17.74 & 95.95 & 130.67 & 79.50 & 15.64 & 153.76 & 0.530 & -0.120 & 65.50 & -9.62 & 0.19 & -5.15 & -8.13 \\
  -0.4 & -64.07 & 46.62 & -15.75 & 86.53 & 154.26 & 66.73 & 23.18 & 169.67 & 0.432 & -0.106 & 52.73 & -10.21 & 0.37 & -5.61 & -8.53 \\
  -0.3 & -47.21 & 52.65 & -13.05 & 76.74 & 174.83 & 50.64 & 29.43 & 184.38 & 0.318 & -0.088 & 36.64 & -10.72 & 0.53 & -6.02 & -8.88 \\
  -0.2 & -28.46 & 56.86 & -9.81 & 67.78 & 190.94 & 31.33 & 33.53 & 196.39 & 0.192 & -0.066 & 17.33 & -11.12 & 0.63 & -6.32 & -9.17 \\
  -0.1 & -8.36 & 58.96 & -6.29 & 61.28 & 200.97 & 9.31 & 34.67 & 204.15 & 0.056 & -0.042 & -4.69 & -11.37 & 0.66 & -6.48 & -9.36 \\
  0.0 & 12.38 & 58.72 & -2.84 & 59.00 & 203.39 & -14.22 & 32.29 & 206.43 & -0.083 & -0.019 & -28.22 & -11.43 & 0.60 & -6.47 & -9.44 \\
  0.1 & 32.96 & 56.08 & 0.20 & 61.69 & 197.76 & -37.02 & 26.69 & 202.97 & -0.222 & 0.001 & -51.02 & -11.29 & 0.46 & -6.29 & -9.38 \\
  0.2 & 52.58 & 51.23 & 2.54 & 68.40 & 184.95 & -57.16 & 18.81 & 194.50 & -0.354 & 0.017 & -71.16 & -10.97 & 0.26 & -5.96 & -9.21 \\
  0.3 & 70.60 & 44.51 & 4.01 & 77.36 & 166.59 & -73.79 & 9.82 & 182.47 & -0.476 & 0.027 & -87.79 & -10.51 & 0.04 & -5.54 & -8.93 \\
  0.4 & 86.53 & 36.27 & 4.55 & 87.00 & 144.30 & -86.71 & 0.72 & 168.35 & -0.583 & 0.031 & -100.71 & -9.96 & -0.18 & -5.06 & -8.58 \\
  0.5 & 100.02 & 26.90 & 4.18 & 96.24 & 119.43 & -96.01 & -7.73 & 153.44 & -0.674 & 0.028 & -110.01 & -9.34 & -0.39 & -4.56 & -8.16 \\
  0.6 & 110.89 & 16.75 & 3.01 & 104.44 & 92.88 & -102.02 & -15.07 & 138.80 & -0.747 & 0.020 & -116.02 & -8.68 & -0.58 & -4.07 & -7.69 \\
  0.7 & 118.97 & 6.13 & 1.15 & 111.16 & 65.07 & -105.05 & -21.06 & 125.37 & -0.802 & 0.008 & -119.05 & -7.99 & -0.72 & -3.58 & -7.18 \\
  0.8 & 124.20 & -4.64 & -1.24 & 116.15 & 37.04 & -105.28 & -25.40 & 114.46 & -0.837 & -0.008 & -119.28 & -7.30 & -0.83 & -3.12 & -6.65 \\
  0.9 & 126.55 & -15.31 & -3.99 & 119.30 & 9.20 & -102.93 & -28.00 & 107.09 & -0.853 & -0.027 & -116.93 & -6.61 & -0.90 & -2.70 & -6.09 \\
  1.0 & 126.09 & -25.61 & -6.91 & 120.56 & -18.26 & -98.20 & -28.84 & 103.99 & -0.850 & -0.047 & -112.20 & -5.93 & -0.92 & -2.33 & -5.53 \\
  1.1 & 122.84 & -35.31 & -9.83 & 119.92 & -45.19 & -91.16 & -27.95 & 105.54 & -0.828 & -0.066 & -105.16 & -5.26 & -0.90 & -2.00 & -4.95 \\
  1.2 & 116.87 & -44.18 & -12.57 & 117.38 & -71.48 & -81.82 & -25.33 & 111.56 & -0.788 & -0.085 & -95.82 & -4.61 & -0.83 & -1.71 & -4.36 \\
  1.3 & 108.25 & -51.97 & -14.95 & 112.97 & -96.87 & -70.13 & -21.00 & 121.44 & -0.730 & -0.101 & -84.13 & -3.98 & -0.72 & -1.47 & -3.76 \\
  1.4 & 97.09 & -58.44 & -16.81 & 106.79 & -121.08 & -55.99 & -15.05 & 134.26 & -0.654 & -0.113 & -69.99 & -3.37 & -0.58 & -1.28 & -3.18 \\
  1.5 & 83.56 & -63.34 & -17.99 & 99.02 & -143.15 & -39.64 & -7.77 & 148.75 & -0.563 & -0.121 & -53.64 & -2.83 & -0.39 & -1.14 & -2.62 \\
  1.6 & 67.91 & -66.47 & -18.36 & 90.05 & -162.44 & -21.11 & 0.55 & 163.81 & -0.458 & -0.124 & -35.11 & -2.35 & -0.19 & -1.07 & -2.10 \\
  1.7 & 50.46 & -67.59 & -17.85 & 80.43 & -178.20 & -0.36 & 9.52 & 178.46 & -0.340 & -0.120 & -14.36 & -1.96 & 0.03 & -1.05 & -1.65 \\
  1.8 & 31.62 & -66.48 & -16.42 & 71.12 & -189.25 & 22.23 & 18.44 & 191.45 & -0.213 & -0.111 & 8.23 & -1.68 & 0.26 & -1.10 & -1.30 \\
  1.9 & 11.96 & -63.00 & -14.11 & 63.54 & -194.22 & 45.90 & 26.38 & 201.32 & -0.081 & -0.095 & 31.90 & -1.56 & 0.45 & -1.20 & -1.10 \\
  2.0 & -7.85 & -57.10 & -11.09 & 59.49 & -191.85 & 69.40 & 32.20 & 206.54 & 0.053 & -0.075 & 55.40 & -1.62 & 0.60 & -1.35 & -1.07 \\
  2.1 & -27.01 & -48.88 & -7.64 & 60.31 & -181.62 & 90.66 & 34.84 & 205.96 & 0.182 & -0.051 & 76.66 & -1.87 & 0.66 & -1.54 & -1.25 \\
  2.2 & -44.76 & -38.70 & -4.09 & 65.73 & -164.65 & 107.64 & 33.93 & 199.63 & 0.302 & -0.028 & 93.64 & -2.29 & 0.64 & -1.74 & -1.63 \\
  2.3 & -60.52 & -27.04 & -0.80 & 74.17 & -142.88 & 119.52 & 29.93 & 188.67 & 0.408 & -0.005 & 105.52 & -2.83 & 0.54 & -1.93 & -2.14 \\
  2.4 & -73.89 & -14.43 & 1.96 & 83.88 & -118.09 & 126.44 & 23.69 & 174.62 & 0.498 & 0.013 & 112.44 & -3.45 & 0.39 & -2.12 & -2.75 \\
  2.5 & -84.62 & -1.34 & 4.00 & 93.54 & -91.76 & 128.87 & 16.11 & 159.03 & 0.570 & 0.027 & 114.87 & -4.10 & 0.20 & -2.30 & -3.40 \\
  \hline
\end{tabular}\\
$t$ is listed in $\myr$, $\{x,y,z,R\}$ in $\pc$, $\{v_x,v_y,v_z,v_{\rm orb},v_{\rm los}'\}$ in $\kms$, $\{l,b\}$ in degrees, and $\{\mu_l',\mu_b',\mu_x',\mu_y'\}$ in ${\rm mas}~\yr^{-1}$.
\end{minipage}
\end{table*}
Table~\ref{tab:fullorbit} shows the complete solution of the best-fitting orbit at time intervals of $\Delta t=0.1~\myr$. The data are also available in machine-readable format in the Supporting Information accompanying this paper, where we use time intervals of $\Delta t=0.01~\myr$.

The first nine columns of Table~\ref{tab:fullorbit} list the main model quantities. Column~1 shows the time $t$, where $t=0$ corresponds to the pericentre passage on the far side of the stream, between Streams~3 and~4 (cf.~Fig.~\ref{fig:orbit_td}). Columns~2--4 give the spatial coordinates $\{x,y,z\}$, where $z=0$ corresponds to the Galactic plane (i.e.~$b=0^\circ$) and $x=0$ corresponds to the $l=0^\circ$ meridian. The $y$-coordinate indicates the distance along the line of sight. Note that $x$ increases to the right (i.e.~towards negative $l$), $y$ increases away from the observer, and $z$ increases to the top (i.e.~towards positive $b$). In these coordinates, the position of the bottom of the gravitational potential at Sgr~A$^*$ is $\{x,y,z\}_{{\rm Sgr~A}^*}=\{8.08,0,-6.68\}~\pc$. Columns~6--8 provide the velocities along these coordinate axes $\{v_x,v_y,v_z\}$. Finally, Columns~5 and~9 provide the total galactocentric radius (defined as $R^2=x^2+y^2+z^2$) and orbital velocity (defined as $v_{\rm orb}^2=v_x^2+v_y^2+v_z^2$).

The remainder of Table~\ref{tab:fullorbit} lists observable quantities. Note that the numbers listed in columns~12--16 include the solar motion to best reflect directly observable quantities (see below). Columns~10 and~11 provide the orbital structure in the plane of the sky, in the Galactic coordinates $\{l,b\}$. The conversion to angular coordinates assumes a distance to the Galactic Centre of $R=8.3~\kpc$ \citep{reid14}. The line-of-sight velocity $v_{\rm los}'$ is given in column~12. The prime indicates that this is the observable velocity in the local standard of rest, adding the Sun's radial velocity of $U_\odot=14~\kms$ towards the Galactic Centre \citep{schoenrich12} to the modelled line-of-sight velocity as $v_{\rm los}'\equiv v_{\rm los}-U_\odot$.\footnote{If another value of $U_\odot$ is preferred, $U_\odot=14~\kms$ must first be added to the listed values before subtracting the preferred radial velocity towards the Galactic Centre.} The proper motions in Galactic coordinates $\{\mu_l',\mu_b'\}$ implied by the best-fitting orbit are provided in columns~13 and~14. Again, the primes indicate that these proper motions include the proper motion induced by the Sun's orbital motion $\{\mu_l,\mu_b\}_\odot=\{-6.379,-0.202\}~{\rm mas}~\yr^{-1}$ \citep{reid04,reid09} as $\{\mu_l',\mu_b'\}\equiv\{\mu_l,\mu_b\}+\{\mu_l,\mu_b\}_\odot$.\footnote{If another value of $\{\mu_l,\mu_b\}_\odot$ is preferred, $\{\mu_l,\mu_b\}_\odot=\{-6.379,-0.202\}~{\rm mas}~\yr^{-1}$ must first be subtracted from the listed values before adding the preferred solar motion.} Columns~15 and~16 list the proper motions in equatorial coordinates $\{\mu_\alpha',\mu_\delta'\}$.\footnote{This again includes the proper motion induced by the solar motion.}

\bsp

\label{lastpage}

\end{document}